\documentclass[12pt]{iopart}

\usepackage{graphics}
\usepackage{iopams}
\usepackage{fancyhdr}

\newcommand{\torque}{{\tau}}
\newcommand{\bend}{{\Omega}}
\renewcommand{\vec}[1]{\mbox{\boldmath$#1$}}

\newcommand{\uc}{\xi}

\begin{document}

\bibliographystyle{unsrt}

\title[Descriptions of membrane mechanics\dots]{Descriptions of membrane mechanics from microscopic and effective two-dimensional perspectives}

\author{Michael A. Lomholt$^{1,2}$ and Ling Miao$^1$}
\address{$^1$ The MEMPHYS Center for Biomembrane Physics, Physics Department,
University of Southern Denmark, Campusvej 55, DK-5230 Odense M, Denmark}
\address{$^2$ NORDITA - Nordic Institute for Theoretical Physics, Blegdamsvej 17, 2100 Copenhagen \O, Denmark}
\ead{mlomholt@memphys.sdu.dk}

\date{\today}

\begin{abstract}
Mechanics of fluid membranes may be described in terms of the
concepts of mechanical deformations and stresses, or in terms
of mechanical free-energy functions.  In this paper, each of
the two descriptions is developed by viewing a membrane from
two perspectives: a microscopic perspective, in which the membrane
appears as a thin layer of finite thickness and with highly
inhomogeneous material and force distributions in its transverse
direction, and an effective, two-dimensional perspective, in which
the membrane is treated as an infinitely thin surface, with effective
material and mechanical properties.  A connection between these
two perspectives is then established.  Moreover, the functional
dependence of the variation in the mechanical free energy of the
membrane on its mechanical deformations is first studied in the
microscopic perspective.  The result is then used to examine to
what extent different, effective mechanical stresses and forces
can be derived from a given, effective functional of the mechanical
free energy.
\end{abstract}


\maketitle

\section{Introduction} \noindent
Mechanics of fluid, lipid-bilayer based membranes has been one of the most important topics
in membrane physics during the past three decades \cite{evans80,lipowsky95}.  It contains much
richer physics than that of conventional fluid-fluid interfaces, due to the fact that the concept
of surface tensions is, under most physically relevant situations, not sufficient to describe
mechanical states of the membranes, unlike in the case of conventional fluid-fluid interfaces.
Additional mechanical quantities to describe their resistance towards bending
deformation have to be included in the description as well \cite{helfrich73,evans74}.
Understanding of the equilibrium shapes of red-blood cells \cite{wortis02} and giant, artificial
lipid-bilayer vesicles \cite{helfrich73,miao91,miao94,seifert97} relies on a well-defined description
of the mechanical properties of the relevant membranes, so do the applications of membrane
mechanics in experimental techniques such as the micro-mechanical manipulation technique 
\cite{evans80b,henriksen04} and the sensitive force-measurement technique based on the mechanics of
cells or vesicles \cite{evans95,merkel98,merkel99}.  Recently, descriptions of membrane mechanics
have also been employed in the investigation of surface-induced forces between membrane-adsorbed
or membrane-embedded colloidal particles \cite{goulian93,guven05}.    
 
Descriptions of mechanics of fluid membranes are most often formulated from the point of view of
elastic shell theory, or an effective, two dimensional perspective, where the membrane is treated as
an infinitely thin surface. A description is given either in terms of surface stresses
and bending moments, expressed as functions of the deformations, or in terms of mechanical work
or free energy associated with the deformations \cite{evans80,kralchevsky94}.  A state of mechanical
equilibrium of the membrane, or its equilibrium shape, can then be determined by setting up equations
of balance for the stresses and for the bending moments, if they are given.  Obviously
those equations correspond to the conservations of linear momentum and angular momentum,
respectively \cite{evans80,kralchevsky94}.   Alternatively,  the problem of finding equilibrium
shapes of the membrane can be approached, more directly, by applying a variational principle
to derive the equations of mechanical equilibrium from the mechanical free energy
\cite{zhongcan89,miao91,julicher94,seifert97}.  Previous works have shown that these two approaches
can be reconciled with each other \cite{powers02,capovilla02}.  In particular, the work presented
in \cite{capovilla02} based on the use of N{\"o}ther's theorem provides a very general, albeit
somewhat abstract, argument regarding why the two approaches are equivalent.

\begin{figure}
\vspace*{.5cm}
\centerline{
\resizebox{8cm}{!}{
  \includegraphics{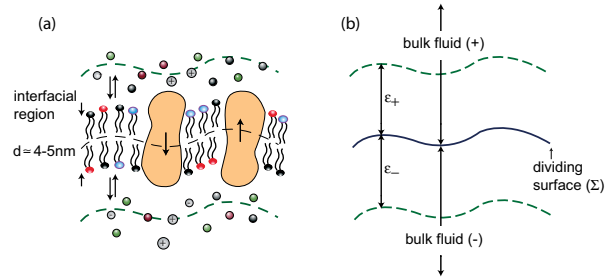}  
}}
\caption{\label{fig:one}A schematic sketch of the membrane-fluid system under our consideration.
Part (a) depicts the interfacial region in the real system, showing a membrane composed of
bilayer-forming lipids and transmembrane proteins in contact with two fluids containing small
solutes. Part (b) illustrates the representation of the interfacial region in the corresponding
Gibbs model system.}
\end{figure}

One of the goals of this paper is to give a more mechanical -- therefore, more experimentally
relevant -- account than that given in \cite{capovilla02}, of why the two approaches of obtaining
equations of membrane mechanical equilibrium are equivalent.  For this purpose, general
descriptions of the mechanics and thermodynamics of a fluid membrane will be developed by viewing the membrane
from two different perspectives.  One is a microscopically realistic perspective, depicted
in Figure \ref{fig:one}(a), where the membrane has a highly inhomogeneous distribution of matter
in its transverse direction.  This will often be referred to as the microscopically viewed
system or the microscopic model.  The other perspective is an effective, or idealized,
one, illustrated in Figure \ref{fig:one}(b), where the membrane is modeled as an infinitely thin
surface, which will be called the dividing surface, with the bulk solvent filling all of
the space on the two sides of the surface.  This model will also be called the Gibbs system
or Gibbs picture, as effective, excess mechanical quantities will be assigned to this surface
following the idea of Gibbs' \cite{gibbs61}.  Those excess quantities are defined by the
constraints of mechanical equivalence between the two different perspectives. In other
words, the total mechanical forces and torques obtained from integrating the linear
stress and the bending moment over a surface which traverses the membrane surface and
has a sufficiently large transverse dimension must be the same in the two perspectives.
If the transverse extensions of the surface of integration into the two solvent regions
on the two sides of the membrane are denoted $\epsilon^+$ and $\epsilon^-$, the term
``sufficiently large transverse dimension" refers to the following requirement:
that $\epsilon^+$ and $\epsilon^-$ be judiciously chosen such that at these distances
from the dividing surface the thermomechanical properties of the solvent must be
indistinguishable from those of a homogeneous bulk solvent under the same thermodynamic
conditions.  

A second goal of the paper is to add to the canonical descriptions
\cite{evans80,kralchevsky94,capovilla02} descriptions of physical situations where applied
external forces that act, or do work, on membranes are not localized on the corresponding
dividing surfaces.  An experimentally relevant example of such situations is the following \cite{harbich79,mitov93}: an external electric
field is applied to a membrane system, induces a dipole moment in the membrane, and through
its coupling to the dipole moment exerts a mechanical torque on the membrane.  The work of
such a torque will depend on membrane deformations and therefore, contribute to the mechanics
of the membrane.  Such situations have not been considered in the earlier works
\cite{evans80,kralchevsky94,capovilla02}. 

A third goal of the paper is to address an issue that arises from the exercise of deriving
mechanical stresses from a given thermomechanical free-energy functional.  It turns out that
mechanical stresses obtained by this way inevitably contain in their expressions some degree of
arbitrariness.  A similar problem has a long history in classical field theory \cite{gotay92}.
In this paper, the origin of the arbitrariness is investigated and the issue of to what extent
this arbitrariness can be eliminated in physically meaningful or relevant ways is discussed.
This part of our study should facilitate the task of determining states of mechanical
equilibrium of membranes.

It should be noted that even though we use the word membrane throughout the paper, then the theory is general enough also to apply to for instance the interface of two coexisting fluids. However, since a membrane is not necessarily situated at such an interface, it can for instance be freely floating, and since the prime example that we have in mind is a lipid-bilayer based membrane, we will use the word membrane throughout the text.

The outline of the paper will be as follows. In Section~\ref{sec:memmech} the Gibbs description of the mechanics of fluid membranes will
be introduced.  In Section \ref{sec:memmicro} a description based on the corresponding
microscopic view of the membrane systems is established and is then connected to the Gibbs
description.  This connection is then used in Section \ref{sec:thermodyn} to elucidate how
the mechanical stresses in the Gibbs description are related to the functional variation
of a general mechanical free energy under shape deformations and to identify the issue
that the mechanical stresses derived from the free-energy variation are not uniquely
determined.  This issue is then addressed in Section~\ref{sec:arbit}.  A conclusion
is given in Section~\ref{sec:concl}, and it is finally followed by two appendices. The first of these, Appendix \ref{sec:geom}, is a brief summary of the differential geometry of surfaces. It is included mostly to have a place to consult for the notation and conventions used.

\section{The Gibbs description of membrane mechanics}\label{sec:memmech}
In this section we will develop a description of membrane mechanics based on viewing 
from an effective, or idealized, perspective a system of a membrane together with the
fluids that surround it.  In this perspective, the whole system is approximated by
a dividing surface of zero thickness together with two regions of bulk fluids, the
behavior of which are assumed not to be affected by the presence of the membrane. We refer to a membrane system modeled in this way as the Gibbs model or the Gibbs system.
The mechanics of the Gibbs system is then described by a set of effective surface
stresses -- defined on the dividing surface only -- together with those known of
the bulk fluids.  We will first focus on developing the concept of effective surface
stresses, and then give a full description of the mechanics of the Gibbs system.
In correspondence with laws of conservation of linear and angular momenta fundamental
to mechanics, we will classify mechanical stresses into linear stress tensors and
angular stress tensors.  We will develop the concept of the linear stress tensors
from the more fundamental concept of forces, and similarly, the concept of the angular
stress tensors from the concept of torques.

\subsection{Forces and linear stresses}
Similar to mechanical stress tensors in conventional material systems, linear stresses in
a membrane are employed to describe the forces acting on an element of the membrane by the
rest of the membrane with which it is in contact.  The nature of such contact forces is that
they are boundary forces.  Denoting the element of the membrane under consideration $\Sigma$,
we can define a vector, $\vec{T}_{(\vec{\nu})}=\vec{T}_{(\vec{\nu})}(\uc^1,\uc^2)$, to be the
force per unit length acting on the boundary of $\Sigma$.   The vector $\vec{\nu}=\nu^\alpha \vec{t}_\alpha$
represents the unit vector which points in the outward normal direction to the boundary element
local to point $(\uc^1,\uc^2)$ and which lies tangentially in the membrane plane.  Clearly,
the total force on $\Sigma$ arising from the boundary forces, $\vec{F}_{\rm \Sigma}$, is then
given by integrating $\vec{T}_{(\vec{\nu})}$ along the boundary of $\Sigma$,
\begin{equation}
\vec{F}_{\rm \Sigma}=\int_{\partial\Sigma} ds\ \vec{T}_{(\vec{\nu})}\ ,
\end{equation}
where $ds$ is the arc length along the boundary $\partial\Sigma$.

Obviously, $\vec{T}_{(\vec{\nu})}$ depends on the orientation of $\vec{\nu}$.  This dependence can be shown to be
\begin{equation}\label{eq:nudep}
\vec{T}_{(\vec{\nu})}=\vec{T}^\alpha\nu_\alpha\ ,
\end{equation}
where the two constituting space force vectors, $\vec{T}^\alpha$, which also form surface vectors, are independent of $\vec{\nu}$. We have for completeness given a derivation of Equation~(\ref{eq:nudep}) using a standard argument \cite{naghdi72} in \ref{sec:form}.  Readers willing to accept the result may skip that appendix.

It should be clear that $\vec{T}^\alpha$ describes the non-trivial local distribution of forces, and it will therefore be identified as the linear surface stress (tensor) in the membrane. It will often be used later in the form of its components in a decomposition of the following form:
\begin{equation}
\vec{T}^\alpha\equiv T^{\alpha\beta}\vec{t}_\beta + T_{{\rm n}}^{\alpha}\vec{n}\;,
\end{equation}
where its tangential components, $T^{\alpha\beta}$, constitute a surface tensor of rank 2, and its transverse components, $T_{{\rm n}}^{\alpha}$, form a surface vector.

Using Equation~(\ref{eq:nudep}) we can now rewrite force $\vec{F}_\Sigma$:
\begin{equation}
\vec{F}_{\rm \Sigma} =\int_{\partial\Sigma} ds\ \vec{T}^\alpha\nu_\alpha 
=\int_\Sigma dA\ D_\alpha\vec{T}^\alpha\;,
\end{equation}
where Gauss's law has been used for the second equality sign.  From this we see that 
a surface density of force can be associated with the stress tensor through its 
covariant divergence, 
\begin{equation}
\vec{f} =D_\alpha\vec{T}^\alpha\;.\label{eq:fandT}
\end{equation}
If $\vec{f}$ is non-zero in equilibrium, then an ``external" force acting on the membrane to
balance it is necessarily involved.  Such an external force can, for example, result from
the stresses exerted on the membrane by the surrounding bulk fluids.  We will return to
this point in Section \ref{sec:memmicro}.

With the help of the components of the linear stress tensor, we can decompose,
$\vec{T}_{(\vec{\nu})}$, the force per unit length acting on a membrane element through one
of its boundary-curve elements with outward normal $\vec{\nu}$, into three components,
\begin{equation}
\vec{T}_{(\vec{\nu})}= T_{\mu}\vec{\nu}+T_{\rm s}\left(\vec{n}\times\vec{\nu}\right)+Q_{\rm T}\vec{n}\;.
\end{equation}
These three components have direct physical interpretations:
$T_{\mu} \equiv T_{\alpha\beta}\nu^\alpha\nu^\beta$ can be identified as the surface tension
acting on the boundary element, $T_{\rm s}=T_{\alpha\beta}\nu^\alpha\nu_\gamma\varepsilon^{\gamma\beta}$
can be identified as the shear tangential to the boundary, and $Q_{\rm T}=T_{\rm n}^\alpha\nu_\alpha$
is transverse shear normal to the surface.  These concepts have already been introduced
and used in previous seminal work on membrane mechanics \cite{evans80}.

\subsection{Torques and angular stresses}
Following a line of reasoning similar to that sketched in the preceding subsection,
we can also develop the concept of angular stresses from the concept of torques.  
Considering the same local element, $\Sigma$, of the membrane, with its boundary curve 
$\partial \Sigma$, we can define two densities of torque: $\vec{\bend}_{(\vec{\nu})}$,
the torque per unit length acting by the part of the membrane neighbouring to $\Sigma$
on an element of the boundary curve with $\vec{\nu}$ as its outward pointing normal vector,
and $\vec{\tau}_{}$, the total torque per unit area resulted from the torques distributed
along the boundary.  In the rest of the paper, it will be assumed implicitly that any torque
referred to is calculated with respect to the origin of the global coordinate system unless specified otherwise. We have, therefore,   
\begin{equation}
\int_\Sigma d A\;\vec{\tau}_{}=\int_{\partial\Sigma} ds\;\vec{\bend}_{(\vec{\nu})}\;.
\end{equation}
We can also establish the following two equations,
\begin{equation}
\vec{\bend}_{(\vec{\nu})}=\vec{\bend}^\alpha\nu_\alpha\;,
\end{equation}
and
\begin{equation}
\label{eq:tauvsomega}
\vec{\torque}_{}=D_\alpha\vec{\bend}^\alpha \;,
\end{equation}
based on the same arguments that have led to Equation~(\ref{eq:nudep}) and Equation~(\ref{eq:fandT}).
Obviously the space-surface vector quantity, $\vec{\bend}^\alpha$, is the counterpart
of $\vec{T}^\alpha$ in the context of angular momentum, and will therefore be called the
angular stress tensor.  

To make clear the physical content of the angular stress tensor, we decompose it into
two parts.  One is the angular stress generated by the action of the linear stress,
which is given by $\vec{R} \times \vec{T}^\alpha$.  The other, defined by
\begin{equation}
\label{eq:omegaintdef}
\vec{N}^\alpha = \vec{\bend}^\alpha - \vec{R} \times \vec{T}^\alpha \;,
\end{equation}
represents a contribution which is non-trivial.  Its origin lies in the fact that the
real physical system of a fluid membrane is not an infinitely thin surface and that the force
distribution along its transverse dimension is inhomogeneous.  This point will be discussed
more in the next Section.  We will call $\vec{N}^\alpha$ the {\em internal} angular
stress tensor in reference to its origin.   In correspondence with the decomposition of
$\vec{\bend}^\alpha$ we divide the surface density of the torque, $\vec{\tau}$, into
two parts, 
\begin{equation}
\vec{\tau}_{} = \vec{R}\times\vec{f}_{}+\vec{\tau}_{\rm int}\;.
\end{equation}
The first part describes the contribution from the surface density of the force, and
the other part, which can easily be rewritten as
\begin{equation}
\label{eq:tauint}
\vec{\tau}_{\rm int} \equiv \vec{\tau}_{} - \vec{R}\times\vec{f}_{} 
                  = D_\alpha\vec{N}^\alpha + \vec{t}_\alpha\times\vec{T}^\alpha \;,
\end{equation}
will be referred to as the internal torque.

\subsection{Mechanical stresses in the Gibbs system}
Having developed in a systematic way the concepts of surface linear stress tensor and 
angular stress tensor, we can formulate precise expressions of the corresponding
three-dimensional stresses that describe the force and torque distributions across 
the whole Gibbs system consisting of the dividing surface and the bulk fluids.

\begin{figure}
\vspace*{.5cm}
\centerline{
\resizebox{8cm}{!}{
\includegraphics{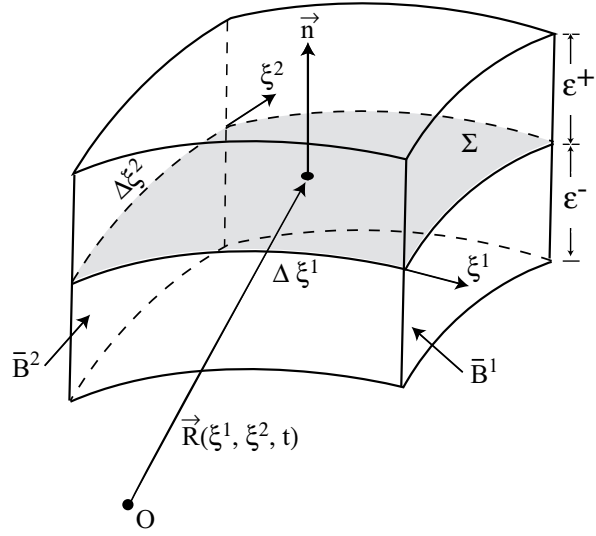}
}}
\caption{A sketch of the cell, or volume element, ${\bar{\Sigma}}$.  $\bar{B}^{1}$ and $\bar{B}^{2}$
refer to side surfaces of the cell. }\label{fig:cell}
\end{figure}

To do that, we consider a three-dimensional cell ${\bar \Sigma}$, as shown in Figure~\ref{fig:cell},
which includes the local element of the dividing surface $\Sigma$, but which has sufficiently large
extension in the transverse direction.  The four side faces of cell ${\bar \Sigma}$, which
are transverse to $\Sigma$ are labeled as $\bar{B}^{1}_{{\pm}}$ and
$\bar{B}^{2}_{{\pm}}$, are parametrized mathematically by
\begin{eqnarray}
\label{eq:belowcond}
{\bar{B}^\alpha_{{\pm}}} &= \Big\{&\vec{R}(\bar{\uc}^1,\bar{\uc}^2)
                     + h \vec{n}(\bar{\uc}^1,\bar{\uc}^2)\;\;\Big|\nonumber\\
                      & & \uc^{3-\alpha}-\Delta\uc^{3-\alpha}/2\leq \bar{\uc}^{3-\alpha}\leq \uc^{3-\alpha} 
		         +\Delta\uc^{3-\alpha}/2\;,\nonumber\\
                      & & \bar{\uc}^\alpha= \uc^\alpha\pm\Delta\uc^\alpha/2\; ,\;\; -\epsilon^-\leq h \leq \epsilon^+\Big\}\;.
\end{eqnarray}

The intersections of $\bar{B}^{\alpha}_{{\pm}}$ with the dividing surface,
${B}^{\alpha}_{{\pm}}$, hence constitute the boundary, $\partial \Sigma$, of
of the dividing surface element $\Sigma$.  

To formulate mathematically concise expressions of the relevant mechanics quantities
of the whole Gibbs system, we introduce two Heaviside step functions, which are defined
as $\theta^\pm(\vec{r})\equiv \theta(\pm \phi(\vec{r}))$.  The scalar function, $\phi(\vec{r})$, appearing in
the definition is defined to be such that it is zero on the dividing surface, positive
in the bulk region that $\vec{n}$ points into and negative in the other.  We also use
two three-dimensional stress tensors, $\mathsf{\bar T}^\pm$, to represent the thermomechanical
properties of the bulk fluids, which are assumed known in terms of their dependences
on thermomechanical control variables. One of the mechanical quantities, which will
be needed soon, is the total force resulted from integrating the three-dimensional
stress tensor over any of the four side surfaces, $\bar{B}^{\alpha}_{{\pm}}$,
\begin{equation}
\label{eq:Gibbs_T}
\vec{F}_{\rm Gibbs}(B^\alpha_{{\pm}}) =
   \int_{B^\alpha_{{\pm}}} d s\ \vec{T}^\beta\nu_\beta  + \int_{{\bar B}^\alpha_{{\pm}}} d{\bar A}\;\vec{\bar \nu}\cdot\left(\theta^+\mathsf{\bar T}^+
  + \theta^-\mathsf{\bar T}^-\right)\;,
\end{equation}
where $\vec{\bar \nu}$ is the outward-pointing unit vector normal to the relevant side surface.

\section{A microscopic description of membrane mechanics}\label{sec:memmicro}
Having introduced different mechanical quantities that constitute the Gibbs description
of mechanics of a fluid membrane, we will in this Section approach the mechanics of
the membrane from the microscopic perspective briefly described in the Introduction.
In this perspective, distributions of forces and torques, or mechanical stress tensors,
are considered three dimensional, and in particular, the inhomogeneity in the distributions
along the dimension transverse to the membrane surface is described explicitly.  Our main
goal is to establish a concrete connection between this microscopic description and
the Gibbs description.  The physical concept that underlies this connection is that of mechanical equivalence. Specifically speaking, we will build up this
connection by considering the linear stresses, the corresponding forces, the angular
stresses, and the corresponding torques one by one.  A similar approach, dealing with the cases of the stresses, can be found for instance
in Ref.~\cite{kralchevsky94}.

\subsection{The linear stress tensors}
In the microscopic description, a linear stress tensor, $\mathsf{\bar T}_{\rm real}$, is
employed to describe the local distribution of force around any point in the three dimensional
space occupied by the membrane and its surrounding fluids.  It is a tensor of rank 2.  To
relate this tensor to the effective linear stress tensor, $\vec{T}^\alpha$, which is defined
only on the two dimensional dividing surface, we apply the concept of mechanical equivalence
explicitly.  It states that in the Gibbs and the microscopic descriptions, the total forces,
resulted from integrating their respective linear stress tensors over each of any two independent
side faces chosen out of $\bar{B}^{\alpha}_{{\pm}}$ must be the same.  In expression,
the statement reads as
\begin{equation}
\label{eq:Gibbsvsreal}
\int_{{\bar B}^\alpha_{{\pm}}} d{\bar A}\ \vec{\bar \nu}\cdot\mathsf{\bar T}_{\rm real} 
= \int_{B^\alpha_{{\pm}}} d s\ \vec{T}^\beta\nu_\beta
+ \int_{{\bar B}^\alpha_{{\pm}}} d{\bar A}\;\vec{\bar \nu}\cdot\left(\theta^+\mathsf{\bar T}^+
+ \theta^-\mathsf{\bar T}^-\right) \;,
\end{equation}
where Equation~(\ref{eq:Gibbs_T}) has been used.

Alternatively, we can define a stress tensor, $\mathsf{\bar T}_{\rm excess}$, by
\begin{equation}
\mathsf{\bar T}_{\rm excess}\equiv\mathsf{\bar T}_{\rm real}-\theta^+\mathsf{\bar T}^+-\theta^-\mathsf{\bar T}^-\;.
\end{equation}
This tensor represents all the excess stress that results from the existence of the membrane interface.
Using it we can rewrite Equation~(\ref{eq:Gibbsvsreal}) as
\begin{equation}
\label{eq:2Dv3D}
\int_{B^\alpha_{{\pm}}} d s\ \vec{T}^\beta\nu_\beta
= \int_{{\bar B}^\alpha_{{\pm}}} d{\bar A}\ \vec{\bar \nu}\cdot\mathsf{\bar T}_{\rm excess}\;.
\end{equation}

To derive from Equation~(\ref{eq:2Dv3D}) an explicit expression of $\vec{T}^\alpha$ in terms
of the microscopic stress tensor we must rewrite the two sides of the Equation~(\ref{eq:2Dv3D}). 
For the rewriting of the right-hand side we choose to work with the curvilinear coordinates
$(\uc^1,\uc^2,h)$, which parametrize the spatial position of any point $\vec{\bar R}$ in the
system by 
\begin{equation}
\label{eq:3Dcoord}
\vec{\bar R}(\uc^1,\uc^2,h) = \vec{R}(\uc^1,\uc^2) + h \vec{n}(\uc^1,\uc^2)\;.
\end{equation}
We then parametrize any of the boundary-curve elements, $B^\alpha_{{\pm}}$, by
$\vec{R}(\uc^1(\lambda),\uc^2(\lambda))$, in a chosen orientation such that the outward pointing
normal vector to the corresponding side face, ${\bar B}^\alpha_{{\pm}}$, is given by
\begin{equation}
\vec{\bar \nu} =\left|\vec{\bar t}_\alpha\frac{d \uc^\alpha}{d\lambda}\right|^{-1}
              \left(\vec{\bar t}_\beta\times\vec{n}\right)\frac{d \uc^\beta}{d\lambda} \;,
\end{equation}
where
\begin{equation}
\vec{\bar t}_\beta\equiv \partial_\beta\vec{\bar R}
=\left(\delta^\gamma_{\beta} - hK^{\gamma}_{\phantom{\gamma}\beta}\right)\vec{t}_\gamma\;.
\end{equation}
It follows then that
\begin{equation}
\label{eq:d_area_vector}
d\bar{A}\,\vec{\bar\nu}=d\lambda dh\,\left(\vec{\bar t}_\beta\times\vec{n}\right)\frac{d \uc^\beta}{d\lambda}\;.
\end{equation}
Using $\vec{t}_\gamma\times \vec{n}=\vec{t}^{\alpha}\varepsilon_{\alpha\gamma}$ we get
\begin{equation}
\int_{{\bar B}^\alpha_{{\pm}}} d{\bar A}\;\vec{\bar \nu}\cdot\mathsf{\bar T}_{\rm excess}=
\int_{{\bar B}^\alpha_{{\pm}}} d\lambda dh\;\vec{t}^{\alpha}\cdot\mathsf{\bar T}_{\rm excess}
\varepsilon_{\alpha\gamma} \left(\delta^\gamma_\beta-h K^\gamma_{\phantom{\gamma}\beta}\right)
\frac{d\uc^\beta}{d\lambda}\;.\label{eq:dhexcess}
\end{equation}
Rewriting the left-hand side of Equation~(\ref{eq:2Dv3D}) by use of the parametrization we get
\begin{eqnarray}
\int_{{B}^\alpha_\pm} ds\ \vec{T}^\alpha\nu_\alpha &=& \int_{{B}^\alpha_\pm} d\lambda\ \vec{T}^\alpha
\left[\vec{t}_\alpha\cdot\left(\vec{t}_\beta\times\vec{n}\right)\right]\frac{d\uc^\beta}{d\lambda}\nonumber\\
&=&\int_{{B}^\alpha_\pm} d\lambda\ \vec{T}^\alpha\varepsilon_{\alpha\beta}\frac{d\uc^\beta}{d\lambda}\;.\label{eq:suk}
\end{eqnarray}
Comparing Equation~(\ref{eq:suk}) with Equation~(\ref{eq:dhexcess}), we finally arrive at the
expression that we need:
\begin{eqnarray}
\label{eq:3Dto2D}
\vec{T}^\alpha(\uc^1,\uc^2) = \int dh&&\left[ g^{\alpha\beta}(\uc^1,\uc^2)
                                      -h L^{\alpha\beta}(\uc^1,\uc^2) \right]\vec{t}_\beta(\uc^1,\uc^2)\nonumber\\
&&\cdot\mathsf{\bar T}_{\rm excess}(\uc^1,\uc^2,h) \;.
\end{eqnarray}

\subsection{The resultant force densities}
$\vec{f}$, the surface density of the force resulted from the surface excess linear stress,
has a counterpart $\vec{\bar f}_{\rm excess}\equiv \vec{\nabla}\cdot\mathsf{\bar T}_{\rm excess}$
in the microscopic description, which is the volume density of the force resulted from the
three-dimensional excess linear stress tensor.  Following the derivation of Equation~(\ref{eq:3Dto2D})
we can now find in a straightforward way an expression of $\vec{f}_{}$ in terms of 
$\vec{\bar f}_{\rm excess}$. 

Using the definition of $\vec{f}$, given in Equation~(\ref{eq:fandT}), together with
Equation~(\ref{eq:2Dv3D}) and applying Gauss' law we have
\begin{equation}
\int_\Sigma d A\;\vec{f}_{}  = \int_{\partial\Sigma} ds\ \vec{T}^\alpha\nu_\alpha=\int_{\partial\bar\Sigma} d{\bar A}\ \vec{\bar \nu}\cdot\mathsf{\bar T}_{\rm excess} = \int_{\bar\Sigma}d{\bar V}\;\vec{\bar f}_{\rm excess} \;.\label{eq:intfrs}
\end{equation}
Working further based on the parametrization defined by Equation~(\ref{eq:3Dcoord}),
we can express the volume element in the above integration as
\begin{equation}
\label{eq:sqrtG}
d{\bar V}=\left(1-2hH+h^2K\right)d A d h\;.
\end{equation}
Since Equation~(\ref{eq:intfrs}) holds for any $\Sigma$ of arbitrarily small area we
conclude then that
\begin{equation}
\label{eq:frsvsfexc}
\vec{f}_{}=\int d h\;\left(1-2hH+h^2K\right)\vec{\bar f}_{\rm excess}\;.
\end{equation}

To illustrate the physical meaning of $\vec{\bar f}_{\rm excess}$ in another way, we
derive for it an alternative expression.  We introduce first the volume density of
the resultant force in the microscopic description of the system, which
is given by
\begin{equation}
\vec{\bar f}_{\rm real} \equiv \vec{\nabla}\cdot\mathsf{\bar T}_{\rm real} \;,
\end{equation}
and those in the bulk fluids of the Gibbs model, which are given by
\begin{equation}
\vec{\bar f}^\pm \equiv \vec{\nabla}\cdot\mathsf{\bar T}^\pm\;.
\end{equation}
Using these quantities we can then express $\vec{\bar f}_{\rm excess}$ at any point
$\vec{r}$ in the system as follows
\begin{equation}
\vec{\bar f}_{\rm excess} =  \vec{\bar f}_{\rm real} - \theta^+\vec{\bar f}^+ - \theta^-\vec{\bar f}^- - \int_{\rm M} d A\;\vec{n}\cdot\left(\mathsf{\bar T}^+
		           - \mathsf{\bar T}^-\right)\delta\left(\vec{r} -\vec{R}\right)\;,\label{eq:fexcform}
\end{equation}
where the subscript ${\rm M}$ indicates that the integral is over all of the membrane and
$\vec{R}$ is the membrane shape function.

Based on Equation~(\ref{eq:fexcform}), it is clear that $\vec{\bar f}_{\rm excess}$ consists
of not only the difference in the volume force densities, $\vec{\bar f}_{\rm real}$ and $\vec{\bar f}^{\pm}$, but also any non-zero
stress difference across the dividing surface,
$\vec{n}\cdot\left(\mathsf{\bar T}^+-\mathsf{\bar T}^-\right)$, associated with the Gibbs
bulk fluids. For example, in the simple case of a conventional system of coexisting fluids separated by an interface, where there is no applied external force, the effective linear stress tensor $\vec{T}^\alpha$ is characterized by a single mechanical quantity, a homogeneous surface tension $\sigma$;
\begin{equation}
\vec{T}^\alpha=\sigma \vec{t}^\alpha
\end{equation}
The effective resultant force $\vec{f}$ is then $2H\sigma \vec{n}$. Equation~(\ref{eq:fexcform}) thus collects all the microscopic effects that contribute to $\vec{f}$. If $\vec{\bar f}_{\rm real}=0$ and $\vec{\bar f}^\pm=0$ are inserted into Equation~(\ref{eq:fexcform}) as the conditions of mechanical equilibrium for the whole system in the microscopic model and for the bulk fluids in the Gibbs model, respectively, Equation~(\ref{eq:fexcform}) becomes the familiar statement of mechanical equilibrium for the dividing surface in the Gibbs model, namely $\vec{f}$ must balance the stress (pressure) difference across the dividing surface. Where there exists a volume distribution of an external
force in the system which acts on membrane molecules but not the bulk fluids, such as one that could be induced by an electric field, $\vec{\bar f}_{\rm real}$ is non-zero and $\vec{\bar f}^{\pm}=0$. $\vec{\bar f}_{\rm excess}$ thus also includes
the effect arising from any interface-related inhomogeneity in $\vec{\bar f}_{\rm real}$. 

\subsection{Micromechanical expression for the angular stress}\label{sub:micbend}
Similar to the linear stress tensor that we have already discussed, the angular stress in
the realistic microscopic description of a membrane system is a second rank tensor,
denoted by $\mathsf{\bar\Omega}_{\rm real}$ in the following.  Although what we develop
in this section can be applied to cases where the systems under considerations have
internal angular momenta, we will limit ourselves to considering those where internal
angular momenta are not relevant.  For such a system, the angular stress tensor is related
to the linear stress tensor in a straightforward way:
\begin{equation}
\label{eq:angular_stress}
\mathsf{\bar\Omega}_{\rm real} = -\mathsf{\bar T}_{\rm real}\times\vec{\bar R}\;,
\end{equation}
where $\mathsf{\bar T}_{\rm real}$ is assumed symmetric, since a symmetric linear stress
tensor can always be constructed \footnote{In fact, Equation~(\ref{eq:angular_stress}), applies
even in the case where a system possesses internal angular momentum, since it is still
possible to construct a symmetric linear stress tensor which takes into account the
internal angular momentum \cite{mclennan66}.}.  The tensor product is defined as
\begin{equation}
\left(\mathsf{\bar T}_{\rm real}\times\vec{\bar R}\right)_{ij} 
=\epsilon_{jkl}\mathsf{\bar T}_{{\rm real},ik}\vec{\bar R}_l\;,
\end{equation}
where the Latin indices range from one to three and indicate the components in a Cartesian
basis $\{\vec{e}_i\}$, and where $\epsilon_{jkl}$ is a third rank tensor, $\epsilon_{123}=1$ and $\epsilon_{jkl}$ is antisymmetric under
any pairwise interchange of indices. Einstein's summation convention
is also implied here for repeated indices.

To establish the microscopic origin of the effective excess angular stress defined on the dividing
surface in the Gibbs description, $\vec{\bend}$, we use $\mathsf{\bar\Omega}_{\rm real}$ as an
intermediate and follow steps that are very similar to those established in the preceding discussion
concerning the linear stress tensors.  The angular stress tensors in the two Gibbs bulk fluids,
$\mathsf{\bar\Omega}^{\pm}$, are related to their linear stress counterparts according to
$\mathsf{\bar\Omega}^{\pm} =\mathsf{\bar T}^{\pm}\times\vec{\bar R}$.  Subtracting these
contributions from $\mathsf{\bar\Omega}_{\rm real}$, we obtain a tensor of excess angular
stress, 
\begin{equation}
\label{eq:omexcess}
\mathsf{\bar\Omega}_{\rm excess} = -\mathsf{\bar T}_{\rm excess}\times\vec{\bar R}\;.
\end{equation}
Going a few steps further, we arrive at the following expression:
\begin{equation}
\int_{B^\alpha_\pm} ds\;\vec{\bend}^\beta\nu_\beta
=\int_{{\bar B}^\alpha_\pm} d{\bar A}\;\vec{\bar R}\times \left(\vec{\bar \nu}\cdot\mathsf{\bar T}_{\rm excess}\right)\;.
\end{equation}
Using the parametrizations given in Equation~(\ref{eq:3Dcoord}) to Equation~(\ref{eq:d_area_vector}),
we can re-express the left and right-hand sides of the above equations, respectively, as
\begin{equation}
\int_{{B}^\alpha_\pm} ds\;\vec{\bend}^\beta\nu_\beta
= \int_{{B}^\alpha_\pm} d\lambda\ \vec{\bend}^\beta\varepsilon_{\beta\gamma}\frac{d\uc^\gamma}{d\lambda}\;,
\end{equation}
and
\begin{eqnarray}
\int_{{\bar B}^\alpha_\pm} d{\bar A}\;\vec{\bar R}\times \left(\vec{\bar \nu}\cdot\mathsf{\bar T}_{\rm excess}\right)
=\int_{{\bar B}^\alpha_\pm} d\lambda\, d h\ &&\left( \vec{R}
 + h\vec{n}\right)\times \left(\vec{t}^{\alpha}\cdot\mathsf{\bar T}_{\rm excess}\right)\nonumber\\
&&   \cdot\varepsilon_{\alpha\gamma}\left(\delta^\gamma_\beta
   -h K^\gamma_{\phantom{\gamma}\beta}\right)\frac{d\uc^\beta}{d\lambda}\;.  
\end{eqnarray}
Comparing these two expressions we arrive at the final expression which reveals the microscopic
origin of $\vec{\bend}$,
\begin{equation}
\vec{\bend}^\alpha = \int d h\;\left(g^{\alpha\beta} -h L^{\alpha\beta}\right)\big[ \vec{R} \times \left( \vec{t}_\beta \cdot \mathsf{\bar T}_{\rm excess} \right)
+ h\vec{n} \times \left( \vec{t}_\beta \cdot \mathsf{\bar T}_{\rm excess} \right) \big]\;.\label{eq:omegamic}
\end{equation}

The two terms in the above expression correspond naturally to two contributions: a contribution
related to the excess linear stress in the Gibbs model,
\begin{equation}
\vec{R}\times \vec{T}^\alpha
= \int dh\ \left(g^{\alpha\beta}
  -h L^{\alpha\beta}\right)\vec{R}\times\left(\vec{t}_\beta\cdot\mathsf{\bar T}_{\rm excess}\right)\;,
\end{equation}
and a contribution giving rise to the internal excess angular stress,
\begin{eqnarray}
\vec{N}^\alpha &=& \int dh\;h\left(g^{\alpha\beta}
                -h L^{\alpha\beta}\right)\vec{n}\times\left(\vec{t}_\beta\cdot\mathsf{\bar T}_{\rm excess}\right)\nonumber\\
             &=& -\int dh\ h\left(g^{\alpha\beta}
	        -h L^{\alpha\beta}\right)\vec{t}_\gamma\varepsilon^{\gamma\delta}\left(\vec{t}_\beta
		 \cdot\mathsf{\bar T}_{\rm excess}\cdot\vec{t}_\delta\right)\;.\label{eq:Nbend}
\end{eqnarray}
From Equation~(\ref{eq:Nbend}), it is clear that the primary contribution to $\vec{N}^\alpha$ (the one that is least suppressed by the microscopic thickness of the membrane) is the first moment of the distribution of the
linear stress tensor in the transverse direction. Following the nomenclature used in the existing
literature on membrane mechanics \cite{evans80}, we will also call $\vec{N}^\alpha$ the bending moment of the membrane.

Equation~(\ref{eq:Nbend}) also implies that $\vec{N}^\alpha\cdot\vec{n}=0$.  This is a consequence
of the assumption we have made in our description of the membrane mechanics: that there is no
bending moment pointing in the normal direction of the membrane surface.  This assumption is
explicitly expressed in Equation~(\ref{eq:3Dcoord}), and is consistent with our empirical
understanding of the fluid characteristics of the membrane in its lateral dimensions.  This
property of $\vec{N}^\alpha\cdot\vec{n}=0$ allows us to define a related vector quantity,
$\vec{M}^\alpha$, which is given by 
\begin{equation}
\label{eq:Mdef}
\vec{N}^\alpha \equiv \vec{n}\times\vec{M}^\alpha\quad\textrm{and} \quad \vec{M}^\alpha\cdot\vec{n}=0\;.
\end{equation}

$\vec{N}^\alpha$ and $\vec{M}^\alpha$ have only tangential components, 
\begin{equation}
\vec{M}^\alpha=M^{\alpha\beta}\vec{t}_\beta\ ,\quad\vec{N}^\alpha= N^{\alpha\beta}\vec{t}_\beta\ .
\end{equation}
Expressed in terms of these components Equation~(\ref{eq:Mdef}) assumes two alternative forms,
\begin{equation}
M^{\alpha\beta} = -N^{\alpha\gamma}\varepsilon_\gamma^{\phantom{\gamma}\beta}\ ,\quad N^{\alpha\beta}
               = M^{\alpha\gamma}\varepsilon_\gamma^{\phantom{\gamma}\beta}\;.
\end{equation}
Thus we are led to the following micromechanical expression for $M^{\alpha\beta}$
\begin{equation}
\label{eq:microbend}
M^{\alpha\beta} = \int dh\ h\left(g^{\alpha\gamma}
                -h L^{\alpha\gamma}\right)\left(\vec{t}_\gamma\cdot\mathsf{\bar T}_{\rm excess}\cdot\vec{t}^\beta\right)\;.
\end{equation}
The simplicity of this expression compared to (\ref{eq:Nbend}) may already demonstrate why
$M^{\alpha\beta}$ is often more convenient to use than $\vec{N}^\alpha$.  It is also physically
more intuitive, readers familiar with Ref.~\cite{evans80} will find that $M^{\alpha\beta}$ are
exactly what are called moment resultants there.  In what follows we will use $\vec{N}^\alpha$
and $M^{\alpha\beta}$ interchangeably, choosing the one that is the more convenient in the case
at hand.

\subsection{Micromechanical expression for the torque}
A micromechanical expression for the effective area density of excess torque defined
in Equation~(\ref{eq:tauvsomega}), $\vec{\tau}$, can also be developed.  In the microscopic
description we can define a volume density of excess torque as the divergence of the excess
angular stress tensor, i.e.,
\begin{equation}
\label{eq:extorque}
\vec{\bar \tau}_{\rm excess}=\vec{\nabla}\cdot\mathsf{\bar \Omega}_{\rm excess} \;.
\end{equation}
Carrying out steps of derivation similar to those leading to Equation~(\ref{eq:frsvsfexc}) for
the effective area density of excess force, we find that $\vec{\tau}$ is related to its counterpart
in the microscopic description as follows:
\begin{equation}
\label{eq:taumicint}
\vec{\tau}_{}=\int d h\;\left(1-2hH+h^2K\right)\vec{\bar \tau}_{\rm excess}\;.
\end{equation}

We can further relate $\vec{\tau}$ to the force distribution $\vec{\bar f}_{\rm excess}$, 
since
\begin{equation}
\label{eq:torque3d}
\vec{\bar \tau}_{\rm excess} =\vec{\bar R}\times\vec{\bar f}_{\rm excess} \;,
\end{equation}
a consequence of the symmetry of the linear stress tensor in the microscopic description. Inserting (\ref{eq:torque3d}) into (\ref{eq:taumicint}) and performing the integration yields
a more revealing expression for $\vec{\tau}$, or more specifically, for the effective internal
torque, $\vec{\tau}_{\rm int}$, defined in Equation~(\ref{eq:tauint}), 
\begin{equation}
\label{eq:taursexp}
\vec{\tau}_{}=\vec{R}\times\vec{f}_{}+\vec{\tau}_{\rm int}\;,
\end{equation}
where
\begin{eqnarray}
\vec{\tau}_{\rm int} &=& \int d h\;\left( 1-2hH+h^2K\right)\left(\vec{\bar R}
                                  -\vec{R}\right) \times \vec{\bar f}_{\rm excess}\nonumber\\
                  &=& \varepsilon^{\alpha\beta}\vec{t}_\beta\left(D_\gamma M^{\gamma}_{\phantom{\gamma}\alpha}
		    -\vec{n}\cdot\vec{T}_\alpha \right)\;.\label{eq:tauspat}
\end{eqnarray}

It is clear from the above expression that only the tangential components of the force
distribution, $\vec{\bar f}_{\rm excess}$, that are not localized on the dividing surface contributes to the effective internal torque.
Under the condition of mechanical equilibrium, the tangential components of
$\vec{\bar f}_{\rm excess}$ are zero, unless there acts a spatially distributed external force
in the directions tangential to the dividing surface.  Consequently, the internal torque
is zero under the same condition.    

A couple of remarks on $\vec{\tau}_{\rm int}$ may be made here.  First, $\vec{\tau}_{\rm int}$
has not been part of the effective description of membrane mechanics developed in the earlier
works \cite{kralchevsky94,capovilla02,evans80}.  There are, however, experimental situations
where $\vec{\tau}_{\rm int}$ is essential.  An example is when an electric field induces and
couples to an electric dipole moment of the membrane \cite{harbich79,mitov93}.  Thus we will
not neglect $\vec{\tau}_{\rm int}$ here.

Secondly, the fact that the component of $\vec{\tau}_{\rm int}$ in the direction normal to
the dividing surface is always zero makes another point clear.  Using this fact on 
Equation~(\ref{eq:tauint}) by setting its normal component to zero leads to the following
result:
\begin{equation}
T^{\alpha\beta}\varepsilon_{\alpha\beta} 
+ M^{\alpha}_{\phantom{\alpha}\gamma}\varepsilon^{\gamma}_{\phantom{\gamma}\beta}
  K^{\beta}_{\phantom{\beta}\alpha} = 0\;.
\end{equation}
This result states that $T^{\alpha\beta}$ is symmetric only if the bending-moment tensor
$M^{\alpha}_{\phantom{\alpha}\beta}$ and the curvature tensor, $K^{\alpha}_{\phantom{\alpha}\beta}$,
commute, or in other words, if both tensors become diagonal simultaneously, in a properly
chosen basis of tangential vectors at every point on the surface.  This condition may be
satisfied in practice in most cases of membrane systems but not in general, although its generality has been assumed assumed
in some of the earlier literature \cite{evans80}.

\section{Free-energy based description of membrane mechanics}
\label{sec:thermodyn}
The descriptions of the mechanics of membranes that we have developed so far are formulated
directly in terms of the linear and angular stresses, or force and torque densities as the
divergences of the corresponding stresses.  However, often in practice, an alternative
formulation is used, given in terms of mechanical free energies as functionals of deformation
of the membranes under consideration.  In such cases, the mechanical stresses are treated
as quantities derived from the free-energy functionals.  In this Section, we will develop,
based on symmetry principles and mathematical argument only, a {\em general} functional form
of the variation of a membrane mechanical free energy associated with an arbitrary variation
in the deformation of the membrane.  We will view the membrane mechanics from the microscopic
perspective, and classify membrane deformation into two categories: one which is of translational
nature, and the other which corresponds to rotations.  From the general functional form of
the free-energy variation we will then identify those variational quantities that are related
to the mechanical stresses and indeed, clarify the correct relations between them.  At the end
of the Section, we will apply the general formulation to a more specific, and frequently used,
example of a mechanical free energy and derive the mechanical quantities such as $\vec{T}^\alpha$,
$\vec{N}^\alpha$. 

\subsection{The mechanical free energy and deformation of a membrane}
In our consideration, we assume that deformations of the membrane occur at constant
temperatures. Correspondingly, we work with a thermodynamic ensemble where the most
relevant free-energy density is given by ${\bar f}={\bar e}-T{\bar s}$, where ${\bar e}$ is
the energy density, $T$ the temperature and $\bar s$ the entropy density.  Following the
already established procedure for defining the excess of various extensive quantities
we define a volume density of the excess of the mechanical free energy by
\begin{equation}
{\bar f}_{\rm excess}(\vec{r}) \equiv {\bar f}_{\rm real}(\vec{r}) 
                                  - \theta^+(\vec{r}){\bar f}^{+}(\vec{r}) - \theta^-(\vec{r}){\bar f}^{-}(\vec{r}) \;,
\end{equation}
where ${\bar f}_{\rm real}(\vec{r})$ is the volume density of the actual free energy
contained in the system and ${\bar f}^{\pm}(\vec{r})$ are defined to be the free-energy
densities associated with the ``filler" bulk fluids in the Gibbs model.

As before, we consider cell $\bar\Sigma$, whose content of the excess free
energy is given by
\begin{equation}
F_{{\bar \Sigma},{\rm excess}} = \int_{\bar\Sigma} d {\bar V}\;{\bar f}_{\rm excess}\;.
\end{equation}
We then define the scalar quantity $f$ to be the excess free energy per unit area associated with the
dividing surface, i.e.
\begin{equation}
\int_\Sigma d A\;f \equiv F_{\bar \Sigma,{\rm excess}}\;.
\end{equation}
Immediately we have the following equivalent expression of $f$,
\begin{equation}
f = \int d h\;\left(1-2hH+h^2K\right){\bar f}_{\rm excess} \;.
\end{equation}

We consider $f$ to be the core element in a free-energy based, phenomenological model
description of the mechanics of a membrane system of interest and take the functional
dependence of $f$ on the relevant thermomechanical variables as our starting point.  In
the discussion that follows, we will limit ourselves to situations where $f$ is a local
function of surface density fields of excess molecular numbers $n_A$,
where $A$ ranges over the different species in the membrane, as well as of
derivatives of the shape field, $\vec{R}$, of the dividing surface.  But, we will
only consider the functional dependence at a generic level, as what is allowed
by symmetry as well as mathematical requirements.  In other words, we will leave
the specific functional behaviour of $f$ to be a degree of freedom to be fixed in
a particular phenomenological model, and only derive consequences from the generic
functional dependence of $f$ on the variables.

A few remarks may be given here to clarify our limiting assumption.  First, the assumption
that only derivatives of $\vec{R}$, not $\vec{R}$ itself, are included in $f$ is a direct
manifestation of our requirement that $f$ be invariant under rigid translations of the
membrane system.  Moreover, we also assume, or require, that $f$ be rotationally invariant.
In the case where external forces such as gravity or an external electric field do work
on the system during its deformation, this requirement of the invariances in turn means
that $f$ represents only the ``internal" contributions, or in other words, that the
contributions associated with the external forces must be considered in addition.  This
requirement is really a matter of choice, made to facilitate the following formulation
in a technical sense, as it will become clear later on.

Secondly, additional thermodynamic variables or fields can be included in $f$, of course.
As long as these fields remain invariant with respect to rigid translations and rotations
of the whole system, their addition does not lead to any qualitative changes in the
arguments that will be developed in the following.  As an example, we may imagine a
situation where the collective orientation of the constituent lipid molecules of the
membrane under consideration becomes relevant \cite{sarasij02}.  A vector field, $\vec{\phi}$,
may then be used to represent the orientational field.  A number of scalar fields which
have the invariances can then be defined from $\vec{\phi}$: its length, its projection
onto the normal direction to the dividing surface and the angle between its tangent
component, $(\vec{\phi}\cdot\vec{t}^\alpha)\;\vec{t}_\alpha$ and the $\uc^1$-axis in the
internal coordinate basis of the local tangent space.

Lastly, the molecular number density fields $n_A$ appearing in $f$ should be understood as
the excess fields defined as
\begin{equation}
\int_\Sigma d A\;n_A  \equiv \int_{\bar\Sigma} d {\bar V}\;{\bar n}_{A,{\rm excess}} \;.
\end{equation}
In other words, the material content of each molecular species in cell $\bar\Sigma$ in
the Gibbs description must be the same as that in the corresponding microscopic description.  

For the purpose of deriving relevant mechanical quantities from the free-energy functional
$f$, the variation of $f$ induced by an arbitrary, infinitesimal deformation $\delta\vec{R}$
in the shape of the dividing surface becomes an essential quantity.  Although we are not
able to evaluate the specific variation without the knowledge of the specific functional
behaviour of $f$, we can already make a statement concerning the general functional structure
of the variation $\delta f$ from a purely mathematical point of view: $\delta f$ associated
with the deformation can always be organized into the following form,
\begin{equation}
\label{eq:varfree}
\frac{1}{\sqrt{g}}\delta_{}\left(\sqrt{g} f\right)= \frac{1}{\sqrt{g}}\left.\delta\left(\sqrt{g} f\right)\right|_{T,\{\sqrt{g}n_A\}}= -\vec{f}_{\rm rs}\cdot\delta\vec{R} + D_\alpha\left(\vec{\hat S}^\alpha\cdot\delta\vec{R}\right)\;.
\end{equation}
In the above equation, the vector quantity $\vec{f}_{\rm rs}$ is regular in the sense that it
does not involve any differential operators, while  
\begin{equation}
\label{eq:S_operator}
\vec{\hat S}^\alpha=\sum_{n=0}^\infty \vec{S}_{(n)}^{\alpha\beta_1\dots\beta_n}D_{\beta_1}\dots D_{\beta_n}\;,
\end{equation}
may involve differential operators.  The fixed $\sqrt{g}n_A$ constraint for the variation
corresponds to the requirement that the excess number of molecules contained in
the surface element $\Sigma$ be conserved for each species under the deformation.

Two points may already be noted here, concerning Equation~(\ref{eq:varfree}).  The first is
that the variation of the free energy is not sufficient to define $\vec{\hat S}^\alpha$
unambiguously.  We will return to this point later, in Section~\ref{sec:arbit}.  The second
is that $\vec{f}_{\rm rs}=D_\alpha\vec{S}^\alpha_{(0)}$.  This follows from the requirement
that the free energy be translationally invariant, or equivalently, that the right-hand side
of Equation~(\ref{eq:varfree}) vanish for constant $\delta\vec{R}$.

Quantities $\vec{f}_{\rm rs}$ and $\vec{S}_{(n)}^{\alpha\beta_1\dots\beta_n}$'s may be related
to the mechanical quantities already introduced and discussed, such as $\vec{f}$, $\vec{\tau}$,
$\vec{T}^\alpha$.  It is the purpose of the rest of this Section to develop the connections.  
In order to do that, we will start from a microscopic point of view and consider the three dimensional cell
$\bar\Sigma$ and an arbitrary, purely mechanical deformation $\delta\vec{\bar R}$ which it
undergoes.  Given $\delta\vec{\bar R}$, we can establish, based on a mathematical point of
view, the following general functional expression for the corresponding variation of the free
energy, $F_{\bar\Sigma,{\rm excess}}$:
\begin{eqnarray}
\delta F_{\bar\Sigma,{\rm excess}}
&= &-\int_{\bar \Sigma} d{\bar V}\;\vec{\bar f}_{\rm excess}\cdot\delta \vec{\bar R}+\int_{\partial {\bar \Sigma}}d {\bar A}\;\vec{\bar \nu}\cdot\mathsf{\hat {\bar T}}\cdot\delta\vec{\bar R}\nonumber\\
  && - \int_{\partial {\bar \Sigma}}d {\bar A}\;\vec{\bar \nu}\cdot\left[\delta\vec{\bar R}-\delta\left(\vec{R}
    + h\vec{n}\right)\right]{\bar f}_{\rm excess} \;.\label{eq:work3d}
\end{eqnarray}
$\mathsf{\hat {\bar T}}$ in the second integral represents a tensor operator with a general
form as
\begin{equation}
\label{eq:T_operator}
\mathsf{\hat {\bar T}}_{i j}=\sum_{n=0}^\infty\mathsf{\bar T}_{(n),i j k_1\dots k_n}\nabla_{k_1}\dots\nabla_{k_n}\;,
\end{equation}
where quantities $\mathsf{\bar T}_{(n),i j k_1\dots k_n}$ are tensors of different ranks. 

Seen from a physical point of view, the meanings of the three terms in Equation~(\ref{eq:work3d})
are almost obvious.  The first term represents the mechanical work done by any non-zero
volume-distributed external force, which balances $\vec{\bar f}_{\rm excess}$.  The second
term describes the work associated with the forces acting on and distributed over the boundary
surface of cell $\bar\Sigma$ by the environment it is in contact with.  The appearance of 
the tensor operator $\mathsf{\hat {\bar T}}$ may, however, have obscured this physical
interpretation and we will comment on this point a bit later.  The presence of the last term
has to do with the following fact: that the geometry of cell $\bar\Sigma$ is defined in such a way that its
side boundary surfaces always remain orthogonal to the dividing surface. Consequently,
there is a free-energy change associated with the movement of the matter across the cell 
boundary induced by $\delta\vec{\bar R}$.

To explain why we have identified the vector quantity in the first term as $\vec{\bar f}_{\rm excess}$,
let's consider a situation where cell deformation and its derivatives vanish at the cell boundary.
The variation in the cell free energy must then equal the work done by any non-zero external
force $\vec{\bar f}_{\rm real,ext}$ under the volume deformation, i.e., 
\begin{equation}
\delta\left.\left(\int_{\bar\Sigma} d{\bar V}\;\bar f_{\rm real}\right)\right|_{\partial\bar\Sigma}
=\int_{\bar\Sigma}d{\bar V}\;\vec{\bar f}_{\rm real,ext}\cdot\delta\vec{\bar R}
=-\int_{\bar\Sigma} d{\bar V}\;\vec{\bar f}_{\rm real}\cdot\delta\vec{\bar R}\;.\label{eq:realvar}
\end{equation}
In a similar fashion, the variations in the free energies, $\bar f^{\pm}$, stored in the
``filler" bulk fluids in the Gibbs description may be expressed as
\begin{eqnarray}
\left.\delta\left(\int_{{\bar\Sigma}^\pm} d{\bar V}\;\bar f^+\right)\right|_{\rm \partial{\bar\Sigma}}
&=& \int_{{\bar\Sigma}^\pm}d{\bar V}\;\mathsf{\bar T}^\pm:\vec{\nabla}\delta\vec{\bar R}\nonumber\\
&=& -\int_{{\bar\Sigma}^\pm} d{\bar V}\;\vec{\bar f}^\pm\cdot\delta\vec{\bar R}\pm\int_\Sigma d A\;\vec{n}\cdot\mathsf{\bar T}^\pm\cdot\delta\vec{\bar R} \left. \right|_{\Sigma}\;,\label{eq:bulkvar}
\end{eqnarray} 
where the colon means that the two tensors are contracted into a scalar and ${\bar\Sigma}^{\pm}$
represents the parts of $\bar\Sigma$ which are, respectively, above and below the dividing
surface $\Sigma$.  Note that Equation~(\ref{eq:bulkvar}) implies that the free energies of the
bulk fluids do not contain terms that vary with the second spatial derivative of the cell
deformation, in contrast to the excess free energy assigned to the dividing surface.

Subtracting Equation~(\ref{eq:bulkvar}) from Equation~(\ref{eq:realvar}) and using 
Equation~(\ref{eq:fexcform}), we find
\begin{equation}
\label{eq:extwork}
\left.\delta F_{\bar \Sigma,{\rm excess}}\right|_{\partial{\bar\Sigma}}
= -\int_{\bar \Sigma} d{\bar V}\;\vec{\bar f}_{\rm excess}\cdot\delta \vec{\bar R}\;.
\end{equation}
This is exactly the contribution described by the first integral in Equation~(\ref{eq:work3d}).

Concerning the second integral in Equation~(\ref{eq:work3d}), the connection between its physical
meaning and its functional structure, or more specifically speaking, the presence of the tensor
operator, $\mathsf{\hat {\bar T}}$, may not appear obvious. A reason for the presence
of $\mathsf{\hat {\bar T}}$ arises from the fact that any given free-energy functional alone does
not lead to a unique identification of the corresponding stress tensor \footnote{Another reason is 
that non-zero $\mathsf{\bar T}_{(n),i j k_1\dots k_n}$ for $n>0$ become necessary if the
orientational degrees of freedom of the membrane under consideration are relevant, as in the
systems of liquid crystals \cite{gennes98}}.  In other words, a kind of ``gauge freedom"
exists for the free-energy based derivation of stress tensors.  The expression of this gauge
freedom necessarily introduces the operator terms contained in Equation~(\ref{eq:T_operator}).
We will make this point clear by arguing that the non-operator term in $\mathsf{\hat {\bar T}}$,
$\mathsf{\bar T}_0$, may be identified with $\mathsf{\bar T}_{\rm excess}$ introduced earlier,
which is assumed to have a well-defined physical interpretation.  The argument may be constructed
as follows.

We first consider a rigid translation of the whole cell $\bar\Sigma$, i.e.,
$\delta\vec{\bar R}=\delta\vec{R} = \vec{C}$, where $\vec{C}$ is a constant vector.  It follows
immediately that
\begin{equation}
0=\delta F_{\bar\Sigma,{\rm excess}}
= -\vec{C}\cdot\int_{\bar \Sigma} d{\bar V}\;\left(\vec{\bar f}_{\rm excess}
   -\vec{\nabla}\cdot\mathsf{\bar T}_{(0)}\right)\label{eq:work3d2} \;.
\end{equation}
Since $\bar\Sigma$ is arbitrary and can also be made arbitrarily small within the limit of a
continuum description, this means that
\begin{equation}
\vec{\bar f}_{\rm excess}-\vec{\nabla}\cdot\mathsf{\bar T}_{(0)}
=\vec{\nabla}\cdot\left(\mathsf{\bar T}_{\rm excess}-\mathsf{\bar T}_{(0)}\right)=0\;.
\end{equation}
We can then conclude based on a standard theorem in calculus that the difference between $\mathsf{\bar T}_{\rm excess}$ and $\mathsf{\bar T}_{(0)}$ is the curl of another tensor $\mathsf{\bar V}$, i.e.
\begin{equation}
\label{eq:excdiff}
\mathsf{\bar T}_{(0),i j} + \epsilon_{i k l}\nabla_k\mathsf{\bar V}_{l j} = \mathsf{\bar T}_{{\rm excess},i j} \;.
\end{equation}
However, as we will see shortly, this difference can be eliminated by a gauge transformation to a new tensor $\mathsf{\bar T}_{(0)}'$. In other words, given the knowledge of $\mathsf{\bar T}_{{\rm excess},i j}$,
we can always choose a gauge transformation accordingly such that 
\begin{equation}
\label{eq:zeroundef}
\mathsf{\bar T}_{(0),i j}'= \mathsf{\bar T}_{(0),i j} - \epsilon_{i k l}\nabla_k\mathsf{\bar V}_{l j}
= \mathsf{\bar T}_{{\rm excess},i j} \;.
\end{equation}
We assume, therefore, that $\mathsf{\bar T}_{(0)}$ is symmetric and identical to $\mathsf{\bar T}_{\rm excess}$ from now on.

The gauge fixing can be performed because the following redefinition of $\mathsf{\hat {\bar T}}$ in
the second integral of Equation~(\ref{eq:work3d}),
\begin{equation}
\label{eq:3darbit_1}
\mathsf{\hat {\bar T}}_{i j}' \delta {\bar R}_j
= \mathsf{\hat {\bar T}}_{(0)i j} \delta {\bar R}_j
 - \epsilon_{i k l}\nabla_k \left( {\bar V}_{l j} \delta {\bar R}_j\right)\;,
\end{equation}
does not change the value of the integral. Note that $\mathsf{\bar T}_{(1),i j}'= \mathsf{\bar T}_{(1),i j}- \epsilon_{i k l}{\bar V}_{l j} \nabla_k \delta {\bar R}_j$ is also modified by the gauge transformation. In this context, we would like to point out also that a more general transformation of
the tensor operator,
\begin{equation}
\label{eq:3darbit_2}
\mathsf{\hat {\bar T}}_{i j}'\delta {\bar R}_j
= \mathsf{\hat {\bar T}}_{i j}\delta {\bar R}_j
 + \epsilon_{i k l}\nabla_k\left({\hat {\bar W}}_{l j}\delta R_j\right)\;,
\end{equation}
where
\begin{equation}
\mathsf{\hat {\bar W}}_{i j}
= \sum_{n=0}^\infty\mathsf{\bar W}_{(n),i j k_1\dots k_n}\nabla_{k_1}\dots\nabla_{k_n} \;,
\end{equation}
is itself another tensor operator, also leaves the free energy unchanged. Equation~(\ref{eq:3darbit_2})
will be used later.

Equation~(\ref{eq:work3d}) will provide the connection between the variational quantities
defined in Equation~(\ref{eq:varfree}) and the mechanical quantities developed in the 
effective (Gibbs) description of the membrane mechanics, once a functional dependence
of the cell deformation $\delta \vec{\bar R}$ on the deformation of the dividing surface,
$\delta \vec{R}$, can be established.  In the following two Subsections, we consider situations
where the cell deformation $\delta\vec{\bar R}$ is a local functional of the deformation of
the dividing surface, $\delta\vec{R}$, together with the shape itself, $\vec{R}$, and possibly other fields like the density fields $n_A$.  We will take the parametrization of a material ``particle" in the membrane-fluid system to be the $\left(\uc^1,\uc^2,h\right)$-coordinates it had before the deformation took place. Given the requirement that the
mechanics of the system be invariant under rigid translations, we may propose a general
form for a description of the cell deformation: 
\begin{eqnarray}
\delta\vec{\bar R} &=& \delta(\vec{R}+h\vec{n})
                  +\mathsf{\hat{\bar \Xi}}^\alpha\cdot\partial_\alpha\delta\vec{R} \nonumber\\
               &=&\delta\vec{R}-h\vec{t}^\alpha(\vec{n}\cdot\partial_\alpha\delta\vec{R})
	          +\mathsf{\hat{\bar \Xi}}^\alpha\cdot\partial_\alpha\delta\vec{R}\;.\label{eq:deform}
\end{eqnarray}
Each $\mathsf{\hat{\bar \Xi}}^\alpha$ for $\alpha = 1,2$ is again a tensor operator.  It should
be clear to the reader that the differential operator $\partial_\alpha$ is included in the above
equation to ensure that $\delta\vec{\bar R} = \delta \vec{R} $ under a rigid translation.   

The specific form of $\mathsf{\hat{\bar \Xi}}^\alpha$ depends on the particular model under
consideration, and a physical example will be given at the end of this Section.  A couple of
general properties of $\mathsf{\hat{\bar \Xi}}^\alpha$ can, however, be readily derived.  The
first is a consequence of the fact that under a rigid, infinitesimal rotation of the system,
$\delta\vec{R}=\vec{\xi}\times\vec{R}$ and $\delta\vec{\bar R}=\vec{\xi}\times\vec{\bar R}$,
where $\vec{\xi}$ is a constant vector.  Applying this requirement to Equation~(\ref{eq:deform})
leads to a condition on $\mathsf{\hat{\bar \Xi}}^\alpha$:
\begin{equation}
\mathsf{\hat{\bar \Xi}}^\alpha\times\vec{t}_\alpha=0\;.
\end{equation}
The second property,
\begin{equation}
\left.\mathsf{\hat{\bar \Xi}}^\alpha\right|_{h=0}=0 \;,
\end{equation}
is an obvious one, an expression of the fact that $\delta\vec{\bar R}$ should be equal to
$\delta\vec{R}$ at the dividing surface where $h=0$. 

\subsection{Translational shape deformations} 
\label{sub:transwork}
In this Subsection, we assume that $\delta\vec{R}$ represents a deformation resulted from
a local translational movement of the dividing surface.  Inserting Equation~(\ref{eq:deform})
into the general functional form of the cell free energy, Equation~(\ref{eq:work3d}) and performing
the integral over $h$, we get the expression of the free energy in terms of surface-related
quantities only, which reads as
\begin{eqnarray}
\delta F_{\bar\Sigma,{\rm excess}}&=& \int_\Sigma d A\;\Big[-\vec{f}_{}\cdot\delta\vec{R}-\left(\vec{n}\times\vec{\tau}_{\rm int}\right)
                    \cdot\vec{t}^\alpha\left(\vec{n}\cdot\partial_\alpha\delta\vec{R}\right)\nonumber\\
&&\phantom{\int_\Sigma d A\;\Big[}-\vec{\hat \Gamma}^\alpha\cdot\partial_\alpha\delta\vec{R}
                              +D_\alpha\left(\vec{T}^\alpha\cdot\delta\vec{R}
			      +\vec{\hat T}_*^\alpha\cdot\delta\vec{R}\right)\Big]\;.\label{eq:microvar}
\end{eqnarray}
Two vector operators, $\vec{\hat \Gamma}^\alpha$ and $\vec{\hat T}^\alpha_*$ are involved in
the above Equation.  $\vec{\hat \Gamma}^\alpha$  is given by
\begin{equation}
\label{eq:Gamma_op}
\vec{\hat \Gamma}^\alpha 
\equiv \int d h\;\left(1-2hH+h^2K\right)\vec{\bar f}_{\rm excess}\cdot\mathsf{\hat{\bar \Xi}}^\alpha\;,
\end{equation}
and is a tensor operator, but contains a regular, non-operator term. 
\begin{equation}
\vec{\hat T}^\alpha_* \equiv 
\sum_{n=1}^\infty\vec{T}_{(n)}^{\alpha\beta_1\dots\beta_n}D_{\beta_1}\dots D_{\beta_n} 
\end{equation}
is a full differential operator which is related to $\hat{\bar{\mathsf{T}}}$. We are not,
however, particularly interested in the operator $\vec{\hat T}^\alpha_*$, and do not, therefore,
provide its explicit expression.

$\vec{\hat \Gamma}^\alpha\cdot\partial\vec{R}$ can always be reorganized into the following form,
\begin{equation}
\label{eq:gamma_ops}
\vec{\hat \Gamma}^\alpha\cdot\partial_\alpha\delta\vec{R}
= \vec{\gamma}^\alpha_{(0)}\cdot\partial_\alpha\delta\vec{R}
  + D_\alpha\left(\vec{\hat \gamma}^\alpha_*\cdot\delta\vec{R}\right) \;,
\end{equation}
which defines a regular vector $\vec{\gamma}^\alpha_{(0)}$ and a vector operator
\begin{equation}
\vec{\hat \gamma}^\alpha_* \equiv 
\sum_{n=1}^\infty\vec{\gamma}_{(n)}^{\alpha\beta_1\dots\beta_n}D_{\beta_1}\dots D_{\beta_n}\;.
\end{equation}
As we will shortly see, these two quantities will add terms to the stress tensors $\vec{T}^\alpha$ and $\vec{\bend}^\alpha$ that has not been included in the previous formalisms \cite{evans80,kralchevsky94,capovilla02}, and they will be discussed further in a
specific example later. Using Equation~(\ref{eq:gamma_ops}) we can rewrite
Equation~(\ref{eq:microvar}) as
\begin{eqnarray}
\delta F_{\bar\Sigma,{\rm excess}}&=&\int_\Sigma d A\;\Big\{ -\left[ \vec{f}_{}-D_\alpha \left( \vec{\gamma}^\alpha_{(0)}
                                                     +\left(\vec{n}\times\vec{\tau}_{\rm int}\right)
						     \cdot\vec{t}^\alpha\,\vec{n}
					       \right) 
		     \right] \cdot\delta\vec{R} \nonumber\\
&&\phantom{\int_\Sigma d A\;\Big[} + D_\alpha \Big[ \left( \vec{T}^\alpha 
                                                  -\vec{\gamma}^\alpha_{(0)}
						  -\left(\vec{n}\times\vec{\tau}_{\rm int} \right)
						  \cdot\vec{t}^\alpha\vec{n}
					     \right) \cdot\delta\vec{R}  \nonumber\\
&&\phantom{\int_\Sigma d A\;\Big[+D_\alpha\Big(} + \left( \vec{\hat T}_*^\alpha 
                                                    -\vec{\hat \gamma}_*^\alpha\right) 
						    \cdot\delta\vec{R}
				      \Big] \Big\}\;.\quad\label{eq:microvar2}
\end{eqnarray}

Comparing the above equation with Equation~(\ref{eq:varfree}) we can relate the excess
mechanical quantities appearing here to the variational quantities $\vec{f}_{\rm rs}$
and $\vec{S}^\alpha_{(0)}$ derived from Equation~(\ref{eq:varfree}).  Immediately we have
\begin{equation}
\label{eq:SvsT}
\vec{T}^\alpha = \vec{S}^\alpha_{(0)} + \vec{\gamma}^\alpha_{(0)}
                    + \left(\vec{n}\times\vec{\tau}_{\rm int}\right)\cdot\vec{t}^\alpha\vec{n} \;,
\end{equation}
and
\begin{equation}
\label{eq:frsvsf}
\vec{f} = \vec{f}_{\rm rs} + D_\alpha\left[ \vec{\gamma}^\alpha_{(0)}
                       + \left(\vec{n}\times\vec{\tau}_{\rm int}\right)\cdot\vec{t}^\alpha\vec{n} \right]\;.
\end{equation}

The above two equations of connection constitute a new result, which has not been discussed
in the existing literature so far.  It is clear from the equations that the functional
form of $f$ alone is not sufficient for deriving the excess linear stress $\vec{T}^\alpha$
and the surface density of the resultant force, $\vec{f}$: In addition we need the knowledge
of $\vec{\gamma}^\alpha_{(0)}$ and $\vec{\tau}_{\rm int}$, which are non-zero in general
when external forces act on the membrane at positions that are not precisely on the dividing surface. Only in the case where there exist no such spatially distributed external force, both $\vec{\gamma}^\alpha_{(0)}$ and $\vec{\tau}_{\rm int}$ vanish.
Consequently, Equation~(\ref{eq:SvsT}) reduces to $\vec{S}^\alpha_{(0)}=\vec{T}^\alpha$,
and we recover a connection used canonically earlier \cite{kralchevsky94,capovilla02}.
Similarly, Equation~(\ref{eq:frsvsf}) reduces to $\vec{f} = \vec{f}_{\rm rs}$.  

It may help to put the discussion in this Section in a context if we briefly mention
an application of this variational approach: the establishment of equations of mechanical
equilibrium for a membrane, or, the ``equilibrium-shape" equations as they are called in
the literature \cite{zhongcan89,miao91,seifert97}.  Once $\vec{T}^\alpha$ and $\vec{f}$ have been
determined based on the derivation of Equation~(\ref{eq:SvsT}) and Equation~(\ref{eq:frsvsf}) both from
a given functional of the membrane excess free energy, $f$, and from given knowledge of
$\vec{\gamma}^\alpha_{(0)}$ and $\vec{\tau}_{\rm int}$, the equation of mechanical equilibrium
that corresponds to force balance can be established from Equation~(\ref{eq:fexcform}) to be
\begin{equation}
\vec{f}+\vec{f}_{\rm ext}+\vec{n}\cdot\left.\left(\bar{\mathsf{T}}^+-\bar{\mathsf{T}}^-\right)\right|_{\vec{r}=\vec{R}}=0\;,
\end{equation}
where
\begin{equation}
\vec{f}_{\rm ext}=\int d h\;\left(1-2hH+h^2K\right)
\left(\vec{\bar f}_{\rm real,ext} - \theta^+\vec{\bar f}^+_{\rm ext} - \theta^-\vec{\bar f}^-_{\rm ext}\right)\;,
\end{equation}
with $\vec{\bar f}_{\rm real,ext}$ being the external force in the microscopic system and $\vec{\bar f}^\pm_{\rm ext}$ the external forces that are assigned to the bulk fluids in the Gibbs system.
 A simple example of such an application
is given, for instance, by a membrane system, the mechanics of which is described entirely
by the Helfrich bending-energy functional \cite{helfrich73} and where there is no external
force applied.  In this case, the relevant equilibrium shape equation is set up by 
setting $\vec{\gamma}^\alpha_{(0)}$ and $\vec{\tau}_{\rm int}$ equal to zero and by
equating the component of $\vec{f}$ normal to the dividing surface to the ``pressure difference"
across the dividing surface \footnote{Note that the term ``pressure difference" should be
understood correctly as that defined in the Gibbs description.}.  The equation resulted
in agrees with that obtained from the variational procedure presented in, for example,
Ref.~\cite{zhongcan89}.

\subsection{Rotational shape deformations} 
\label{sub:rotwork}
The force $\vec{f}$ and the linear stress $\vec{T}^\alpha$ emerge naturally from
Equation~(\ref{eq:microvar2}), which results from the variation in the cell free energy with
respect to a translational shape deformation, $\delta\vec{R}$, of the dividing surface.
In this Section we demonstrate that in a similar fashion, the torque $\vec{\tau}$ and the
angular stress $\vec{\Omega}^\alpha$ will emerge from the variation in the same free energy
associated with a shape deformation that results from infinitesimal local rotations of the
dividing surface.  

An arbitrary rotational deformation of the dividing surface is described by
$\delta\vec{R}=\vec{\xi}\times\vec{R}$, where $\vec{\xi}=\vec{\xi}(\uc^1,\uc^2)$  is a vector
field prescribing the extent of the local rotations.  Substituting this into Equation~(\ref{eq:deform})
leads to the following expression for the whole-cell deformation:
\begin{equation}
\label{eq:rotdeform}
\delta\vec{\bar R} = \vec{\xi}\times\vec{\bar R}
                 -h\vec{t}^\alpha\left(\vec{R}\times\vec{n}\right)\cdot\partial_\alpha\vec{\xi}
		 +\mathsf{\hat{\bar \Xi}}^\alpha\cdot\partial_\alpha\left(\vec{\xi}\times\vec{R}\right) \;.
\end{equation}
We may now replace $\delta\vec{\bar R}$ in Equation~(\ref{eq:work3d}) with the above expression
and derive an effective expression for the excess mechanical free energy contained in the
cell in terms of the surface-related quantities only.

Before doing that, however, we first make a point that will simplify the calculation of the
free-energy variation based on Equation~(\ref{eq:work3d}).  We have argued already that the
mathematical expression of the free-energy variation may involve a non-zero
$\mathsf{\bar T}_{(1),ijk}$ in the tensor operator $\mathsf{\hat{\mathsf{\bar T}}}$.  But this term does
not appear explicitly in the identification of the linear stress and the force.  It turns out
that $\mathsf{\bar T}_{(1),ijk}$ can be neglected in the identification of the angular stress and
the torque also.  To make this point explicitly, let's consider an infinitesimal rigid rotation of
the whole system, i.e., $\delta\vec{\bar R}=\vec{C}\times\vec{\bar R}$, under which the free
energy of the cell remains invariant in the absence of any external force.  Inserting this
expression into Equation~(\ref{eq:work3d}) and using the fact that the linear stress tensor
$\mathsf{\bar T}_{\rm excess}$ is symmetric, we find
\begin{equation}
0=\delta F_{\bar\Sigma,{\rm excess}}
= C_l\int_{\bar \Sigma} d{\bar V}\;\nabla_i\left(\mathsf{\bar T}_{(1),ijk}\epsilon_{jlk}\right) \;.
\end{equation}

Since $\vec{C}$ and $\bar\Sigma$ are arbitrary the divergence in the integrand must vanish.
The same mathematical theorem that led to Equation~(\ref{eq:excdiff}) allows us to conclude
that
\begin{equation}
\mathsf{\bar T}_{(1),ijk}\epsilon_{jlk}= \nabla_j\left(\epsilon_{ijk}\mathsf{\bar V}_{kl}\right)\;,
\end{equation}
where $\mathsf{\bar V}$ is a mathematically well-defined tensor.  Recollecting
Equation~(\ref{eq:3darbit_2}), which expresses the gauge freedom, or the arbitrariness, in the
definition of the tensor operator, we have
\begin{equation}
\mathsf{\bar T}_{(1),i j k}' = \mathsf{\bar T}_{(1),i j k}
                           +\epsilon_{i k l}\mathsf{{\bar W}}_{(0),l j}
			   +\epsilon_{i m l}\nabla_m\mathsf{{\bar W}}_{(1),l j k} \;.
\end{equation}
We can conclude then that an appropriate $\mathsf{{\bar W}}_{(1),l j k}$ can be chosen such that
\begin{equation}\label{eq:T1gauge}
\mathsf{\bar T}_{(1),ijk}'\epsilon_{jlk}=0 \;.
\end{equation}
We will, therefore, assume that $\mathsf{\bar T}_{(1),ijk}\epsilon_{jlk}$ vanishes in what follows.

Having made the above point, we can now substitute Equation~(\ref{eq:rotdeform}) into 
Equation~(\ref{eq:work3d}) and derive the effective expression of $\delta F_{\bar \Sigma,{\rm excess}}$.
Using Equation~(\ref{eq:T1gauge}), performing the $h$-integral and arranging the result in a revealing form, we arrive at
\begin{eqnarray}
\delta F_{\bar \Sigma,{\rm excess}}&=&-\int_{\Sigma} d{A}\;\bigg\{\vec{\tau}_{}-D_\alpha\vec{\omega}^\alpha_{(0)} -D_\alpha \left[ \left(\vec{n}\times\vec{\tau}_{\rm int}\right)
                                                         \cdot\vec{t}^\alpha \left(\vec{R}\times\vec{n}\right)
						    \right]
				  \bigg\}\cdot \vec{\xi} \nonumber\\
&&+\int_{\partial{\Sigma}} d{s}\;\nu_\alpha \bigg[ -\vec{\omega}^\alpha_{(0)} + \vec{\Omega}^\alpha
                                              -\vec{\hat \omega}^\alpha_*
					      +\vec{\hat \Omega}^\alpha_*\nonumber\\
&&\phantom{+\int_{\partial{\Sigma}} d{s}\;\nu_\alpha \bigg[}-\left(\vec{n}\times\vec{\tau}_{\rm int}\right)
                                                            \cdot\vec{t}^\alpha\left(\vec{R}\times\vec{n}\right)
						     \bigg]\cdot\vec{\xi} \;.\label{eq:totalthing}
\end{eqnarray}
Several new quantities are introduced in the above expression.  In particular, the vector quantity,
$\vec{\omega}^\alpha_{(0)}$, and the vector operator, $\vec{\hat \omega}^\alpha_*$, are
defined by the following operation of the vector operator, $\vec{\hat \Gamma}^\alpha$,
defined earlier in Equation~(\ref{eq:Gamma_op}):
\begin{equation}
\vec{\hat \Gamma}^\alpha\cdot\partial_\alpha\left(\vec{\xi}\times\vec{R}\right)
= \vec{\omega}^\alpha_{(0)}\cdot\partial_\alpha\vec{\xi}+D_\alpha\left(\vec{\hat \omega}^\alpha_*\cdot\vec{\xi}\right)\;,
\end{equation} 
with
\begin{equation}
\vec{\hat \omega}^\alpha_*
=\sum_{n=1}^\infty\vec{\omega}_{(n)}^{\alpha\beta_1\dots\beta_n}D_{\beta_1}\dots D_{\beta_n}\;.
\end{equation}
Obviously, $\vec{\omega}^\alpha_{(0)}$ and $\vec{\hat \omega}^\alpha_*$ are related to
$\vec{\gamma}^\alpha_{(0)}$ and $\vec{\hat \gamma}^\alpha_*$ defined earlier in
Equation~(\ref{eq:Gamma_op}), and for $\vec{\omega}^\alpha_{(0)}$ the connection is given by
\begin{equation}
\vec{\omega}^\alpha_{(0)}=\vec{R}\times\vec{\gamma}^\alpha_{(0)}-\vec{\hat \gamma}^\alpha_* \times\vec{R}\;.
\end{equation}
Another vector
operator, $\vec{\Omega}^\alpha_*$, has a general form given by 
\begin{equation}
\vec{\hat \Omega}^\alpha_*
=\sum_{n=1}^\infty\vec{\Omega}_{(n)}^{\alpha\beta_1\dots\beta_n}D_{\beta_1}\dots D_{\beta_n}\;,
\end{equation}
and is related to $\vec{T}^\alpha_*$.

The effective angular stress $\vec{\Omega}^\alpha$ appearing in Equation~(\ref{eq:totalthing}) must now be related to the corresponding
variational quantities associated with the functional of the surface excess free energy $f$. 
In order to do that, we still need to calculate, for the rotational deformation of the dividing
surface, $\delta\vec{R}=\vec{\xi}\times\vec{R}$, the corresponding variation in $f$.  Inserting
the deformation expression into Equation~(\ref{eq:varfree}) and reorganizing the result, we
obtain
\begin{equation}
\label{eq:spatrot2}
\frac{1}{\sqrt{g}}\delta\left(\sqrt{g}f\right)
= -\left(\vec{R}\times\vec{f}_{\rm rs}\right)\cdot\vec{\xi} + D_\alpha\left(\vec{\hat Q}^\alpha\cdot\vec{\xi}\right)\;.
\end{equation}
The vector operator $\vec{\hat Q}^\alpha$ has the general form,
\begin{equation}
\vec{\hat Q}^\alpha=\sum_{n=0}^\infty\vec{Q}_{(n)}^{\alpha\beta_1\dots\beta_n}D_{\beta_1}\dots D_{\beta_n}\;.
\end{equation}
Its first term is given by the following particular expression: 
\begin{equation}
\vec{Q}_{(0)}^{\alpha}=\sum_{n=0}^{\infty}D_{\beta_1}\dots D_{\beta_n}\vec{R} \times \vec{S}_{(n)}^{\alpha\beta_1\dots\beta_n}\;,
\end{equation}
related to the $\vec{S}_{(n)}^{\alpha\beta_1\dots\beta_n}$'s already derived from
Equation~(\ref{eq:varfree}) and Equation~(\ref{eq:S_operator}). Comparing the line integral in Equation~(\ref{eq:totalthing}) with the complete-derivative term in Equation~(\ref{eq:spatrot2})
we find
\begin{equation}
\vec{\Omega}^\alpha=\vec{Q}^\alpha_{(0)}+\vec{\omega}^\alpha_{(0)}+\left(\vec{n}\times\vec{\tau}_{\rm int}\right)\cdot\vec{t}^\alpha\left(\vec{R}\times\vec{n}\right)\;.
\end{equation}

Finally, we can establish, for instance, the connection between the effective bending moments,
$\vec{N}^\alpha$, and the variational quantity, $\vec{Q}_{(0)}^{\alpha}$,
\begin{equation}
\vec{N}^\alpha = \vec{Q}^\alpha_{(0)} +\vec{\omega}^\alpha_{(0)} -\vec{R}\times\left(\vec{S}^\alpha_{(0)}
                +\vec{\gamma}^\alpha_{(0)}\right)\;,
\end{equation}
where Equation~(\ref{eq:omegaintdef}) and Equation~(\ref{eq:SvsT}) have been used.  This equation
may be written into a more revealing form, where $\vec{N}^\alpha$ is divided into two contributions,
\begin{equation}
\label{eq:bending_in_ext}
\vec{N}^\alpha = \vec{N}^\alpha_{\rm rs} - \vec{\hat \gamma}^\alpha_* \times\vec{R}\;.
\end{equation}
The first contribution, $\vec{N}^\alpha_{\rm rs}$, defined as
\begin{equation}
\label{eq:Nrsdef}
\vec{N}^\alpha_{\rm rs} \equiv \vec{Q}^\alpha_{(0)}-\vec{R}\times\vec{S}^\alpha_{(0)}
                     =-\left( \vec{\hat S}^\alpha-\vec{S}^\alpha_{(0)} \right) \times\vec{R} \;,
\end{equation}
can be obtained solely from the free-energy variation.  The second contribution, 
$-\vec{\hat \gamma}^\alpha_* \times\vec{R}$, requires additional knowledge of the cell deformation
and any non-zero external forces.  Again, Equation~(\ref{eq:bending_in_ext}) makes it clear that
a description of the membrane mechanics based on a given free-energy density alone is in general
not sufficient for identifying the effective mechanical quantities of the membrane.

\subsection{Example}
\label{sub:example}
To make the so-far rather general and formalistic discussion more concrete, we end this
Section with an example.  In the example, the surface density of the excess free energy
is a given function of temperature $T$, some number density fields $n_A$'s, the mean curvature
$H$ and the Gaussian curvature $K$ defined for the dividing surface, i.e., $f=f(T,H,K,\{n_A\})$. 
A specific example of such a function may be found in Ref.~\cite{miao02}. A simple purely geometric example is given by the Helfrich free energy \cite{helfrich73}
\begin{equation}
f=\frac{\kappa}{2}\left(2H- C_0\right)^2+{\bar \kappa}K+\sigma\;,\label{eq:helf}
\end{equation}
where $\kappa$, $C_0$, ${\bar \kappa}$ and $\sigma$ are phenomenological constants which are called bending rigidity, spontaneous curvature, Gaussian rigidity and tension respectively.

Following the framework we have already established, we first calculate the variation of $f=f(T,H,K,\{n_A\})$. Using the following set of identities from differential geometry,
\begin{eqnarray}
\delta\left(\sqrt{g}\right) &=& \frac{1}{2}\sqrt{g}g^{\alpha\beta}\delta g_{\alpha\beta}\;,\\
\delta H &=& -\frac{1}{2}K^{\alpha\beta}\delta g_{\alpha\beta}
          +\frac{1}{2}g^{\alpha\beta}\delta K_{\alpha\beta}\;,\\
\delta K &=& -Kg^{\alpha\beta}\delta g_{\alpha\beta}+L^{\alpha\beta}\delta K_{\alpha\beta}\;,
\end{eqnarray}
and carrying out the variation, we obtain
\begin{equation}
\label{eq:metriccurvform}
\frac{1}{\sqrt{g}}\delta\left(\sqrt{g}f\right)
=\frac{1}{2}\sigma^{\alpha\beta}\delta g_{\alpha\beta}-\lambda^{\alpha\beta}\delta K_{\alpha\beta}\;,
\end{equation}
where
\begin{eqnarray}
\sigma^{\alpha\beta} &=& \left( f_{} -2\frac{\partial f_{}}{\partial K}K
                          -\sum_A\frac{\partial f_{}}{\partial n_A}n_A \right)
		     g^{\alpha\beta}-\frac{\partial f_{}}{\partial H}K^{\alpha\beta}\;,\label{eq:sigmaexp}\\
\lambda^{\alpha\beta} &=& -\frac{1}{2}\frac{\partial f_{}}{\partial H} g^{\alpha\beta}
                      -\frac{\partial f_{}}{\partial K}L^{\alpha\beta}\;.\label{eq:lambdaexp}
\end{eqnarray}
Further using
\begin{eqnarray}
\delta g_{\alpha\beta} &=& \vec{t}_\alpha\cdot\partial_\beta\delta\vec{R}
                      +\vec{t}_\beta\cdot\partial_\alpha\delta\vec{R}\ ,\label{eq:deltag2}\\
\delta K_{\alpha\beta} &=& \vec{n}\cdot D_\beta\partial_\alpha\delta \vec{R}\;,\label{eq:deltaK2}
\end{eqnarray}
we can rewrite Equation~(\ref{eq:metriccurvform}) in its final form:
\begin{eqnarray}
\label{eq:shapeform}
\frac{1}{\sqrt{g}}\delta\left(\sqrt{g}f\right) &=& -D_\alpha \left[ \left( \sigma^{\alpha\beta}-\lambda^{\alpha\gamma}K_\gamma^{\phantom{\gamma}\beta}\right)
                     \vec{t}_\beta + D_\beta\lambda^{\alpha\beta}\, \vec{n}
	       \right] \cdot\delta\vec{R}\nonumber\\
&& +D_\alpha \big\{ \big[ \left(\sigma^{\alpha\beta}
                               -\lambda^{\alpha\gamma}K_\gamma^{\phantom{\gamma}\beta}
			 \right) \vec{t}_\beta +D_\beta\lambda^{\alpha\beta} \,\vec{n}\nonumber\\
&&\phantom{ +D_\alpha \big\{ \big[  } -\lambda^{\alpha\beta}\vec{n}\partial_\beta\big]\cdot\delta\vec{R}\big\}\;.
\end{eqnarray}
Comparing the above equation with Equation~(\ref{eq:varfree}) we can read off the various
variational quantities one by one. $\vec{S}^\alpha_{(0)}$, which will contribute to the
linear stress $\vec{T}^\alpha$, is given by
\begin{equation}
\vec{S}^\alpha_{(0)}=\left( \sigma^{\alpha\beta}-\lambda^{\alpha\gamma}K_\gamma^{\phantom{\gamma}\beta}\right)\vec{t}_\beta + D_\beta\lambda^{\alpha\beta}\, \vec{n}\;.\label{eq:S0exp}
\end{equation}
It should be pointed out here that a similar expression was obtained in \cite{guven04} by a sophisticated approach where geometric quantities were treated as auxiliary variables, which were constrained to match the appropriate expressions in terms of the shape field $\vec{R}$ through the Lagrange multiplier formalism. Inserting Eq. (\ref{eq:sigmaexp}) and Eq. (\ref{eq:lambdaexp}) into Eq. (\ref{eq:S0exp}) we arrive at
\begin{eqnarray}
\vec{S}^\alpha_{(0)}\cdot\vec{t}^\beta
%&=& \sigma^{\alpha\beta}-\lambda^{\alpha\gamma}K_{\gamma}^{\phantom{\gamma}\beta} \nonumber\\
&=& \left( f_{}-\frac{\partial f_{}}{\partial K}K-\sum_A\frac{\partial f_{}}{\partial n_A}n_A \right)g^{\alpha\beta}  -\frac{1}{2}\frac{\partial f_{}}{\partial H}K^{\alpha\beta}\;,\label{eq:Ttex}\\
\vec{S}^\alpha_{(0)}\cdot\vec{n}
&=&
% D_\beta\lambda^{\beta\alpha} = 
D_\beta\left( -\frac{1}{2}\frac{\partial f_{}}{\partial H}g^{\alpha\beta}
                -\frac{\partial f_{}}{\partial K}L^{\alpha\beta} \right)\;.\label{eq:Tnex}
\end{eqnarray}
$\vec{f}_{\rm rs}$, which will contribute to the surface density of the deformation-related force,
$\vec{f}$, is given by its components in the normal and the tangential directions,
\begin{eqnarray}
\vec{f}_{\rm rs}\cdot\vec{n}
&=& D_\alpha \left( \vec{S}^\alpha_{(0)}\cdot\vec{n}\right) 
   +\vec{S}^\alpha_{(0)}\cdot\vec{t}^\beta K_{\alpha\beta} \nonumber\\
&=& 2H\left( f_{} -\frac{\partial f_{}}{\partial K}K -\sum_A\frac{\partial f_{}}{\partial n_A}n_A \right)-\frac{\partial f_{}}{\partial H} \left(2H^2-K\right)\nonumber\\
&& -\frac{1}{2}D_\alpha D^\alpha \frac{\partial f_{}}{\partial H} -L^{\alpha\beta}D_\alpha D_\beta \frac{\partial f_{}}{\partial K}\;,
\end{eqnarray}
\begin{eqnarray}
\label{eq:f_tangent}
\vec{f}_{\rm rs}\cdot\vec{t}_\alpha
&=& D_\beta \left(\vec{S}^\beta_{(0)}\cdot\vec{t}_\alpha\right)
   -\vec{S}^\beta_{(0)}\cdot\vec{n} K_{\alpha\beta} \nonumber\\
&=& -\sum_A n_A\partial_\alpha\frac{\partial f}{\partial n_A}
\;.
\end{eqnarray}
The simplicity of Equation~(\ref{eq:f_tangent}) can be traced back to the reparametrization invariance
of the free energy \cite{lomholt05}.

To identify the variational contribution, $\vec{N}^\alpha_{\rm rs}$, to the bending moments, 
we find first from Equation~(\ref{eq:shapeform}) that 
\begin{eqnarray}
\vec{S}^{\alpha\beta\phantom{\dots\beta_n}}_{(1)}&=&-\lambda^{\alpha\beta}\vec{n}\;,\\
\vec{S}^{\alpha\beta_1\dots\beta_n}_{(n)}&=&0\;,\quad \quad \quad \quad n\ge 2\;.
\end{eqnarray}
Substituting these two results into Equation~(\ref{eq:Nrsdef}) immediately leads to
\begin{equation}
\label{eq:varcon}
\vec{N}^\alpha_{\rm rs}
=\vec{n}\times\left(\lambda^{\alpha\beta}\vec{t}_\beta\right)
=\lambda^{\alpha\beta}\varepsilon_\beta^{\phantom{\beta}\gamma}\vec{t}_\gamma\;.
\end{equation}

The formulas above can straightforwardly be applied to the purely geometric Helfrich free energy in Equation (\ref{eq:helf}). The results match the corresponding expressions derived through the N{\"o}ther approach in \cite{capovilla02}.

Note that $\vec{N}^\alpha_{\rm rs}\cdot\vec{n} = 0$ in this case. In general, however, the 
$\vec{N}^\alpha_{\rm rs}$ obtained from the direct variation of a given free-energy density
does not a priori have a zero normal component. On the other hand, the condition $\vec{N}^\alpha\cdot\vec{n}=0$ must be satisfied, as we have already discussed in Section~\ref{sub:micbend}. We will see in the next section that the normal component of $\vec{N}^\alpha_{\rm rs}$ is completely arbitrary, and thus we can always choose a gauge in which it vanishes. The same is true for the $\vec{\hat \gamma}^\alpha_*\times\vec{R}
$ part, and therefore the condition $\vec{N}^\alpha\cdot\vec{n}=0$ can always be satisfied (as it should be).

We have already pointed out in Section~\ref{sec:thermodyn} that a free energy alone is not sufficient in general for the complete determination of the effective mechanical quantities.  Besides $\vec{\tau}_{\rm int}$, at least two additional quantities, $\vec{\gamma}^\alpha_{(0)}$ and $\vec{\omega}^\alpha_{(0)}$, are needed, which bear information on the details of the three-dimensional deformation of the membrane-fluid system.  To illustrate how those quantities can be calculated, we consider an example, where the membrane changes thickness when its shape is deformed. We will later show how this example can be applied to membranes with a certain kind of active proteins incorporated. Mathematically we can model a change of thickness by the three-dimensional deformation
\begin{equation}
\delta\vec{\bar R}=\delta\left(\vec{R}+h\vec{n}\right)+\delta{\bar h}\vec{n}\;,
\end{equation}
where the change in the transverse direction will be written as an expansion to second order in $h$
\begin{equation}
\delta{\bar h}=-2\left(\zeta_1+\zeta_2 h H\right) h\frac{\delta\sqrt{g}}{\sqrt{g}}+2\zeta_3 h^2\delta H\;.\label{eq:gendeform}
\end{equation}
$\zeta_1$, $\zeta_2$ and $\zeta_3$ are three dimensionless phenomenological parameters.  The term involving $\zeta_1+\zeta_2 h H$ describes the effect of a mechanism where the membrane thickness shrinks
(if $\zeta_1+\zeta_2 h H>0$) when the area of the membrane expands.  The factor of 2 is chosen in order to
get agreement with the definition of the Poisson ratio in elasticity theory \cite{landau86}. 
The term characterized by parameter $\zeta_3$ models a type of
structural changes in the system when the membrane is bend. An example of the effects
which could be described by such a term is the flexoelectric effect \cite{petrov99}.

Comparing the above equation with the equation of definition for the tensor operator
$\mathsf{\hat \Xi}^\alpha$, (\ref{eq:deform}), we find first that it is given by 
\begin{equation}
\mathsf{\hat \Xi}^\alpha= -2 \left(\zeta_1+\zeta_2 h H\right)h \vec{n}\vec{t}^\alpha -2\zeta_3 h^2 K^{\alpha\beta}\vec{n}\vec{t}_\beta
  +\zeta_3 h^2 \vec{n}\vec{n}D^\alpha\;.
\end{equation}
Going through the steps prescribed in Equation~(\ref{eq:Gamma_op}) and Equation~(\ref{eq:gamma_ops}),
we arrive at the following identification of $\vec{\gamma}^\alpha_{(0)}$ and $\vec{\omega}^\alpha_{(0)}$:
\begin{equation}
\vec{\gamma}^\alpha_{(0)}
= -2\left(\zeta_1 p_{\rm excess}+\zeta_2 H Q_{\rm excess}\right)\vec{t}^\alpha -\zeta_3 Q_{\rm excess} K^{\alpha\beta}\vec{t}_\beta
  -\zeta_3 D^\alpha Q_{\rm excess}\,\vec{n}\;,\label{eq:gammaspec}
\end{equation}
and
\begin{equation}
\vec{\omega}^\alpha_{(0)}
= \vec{R}\times\vec{\gamma}^\alpha_{(0)} -\zeta_3 Q_{\rm excess}\varepsilon^{\alpha\beta}\vec{t}_\beta\;,
\end{equation}
where
\begin{eqnarray}
p_{\rm excess} &=&\int d h\;h\left(1-2h H+h^2K\right)\vec{n}\cdot\vec{\bar f}_{\rm excess}\;,\\
Q_{\rm excess} &=&\int d h\;h^2\left(1-2h H+h^2K\right)\vec{n}\cdot\vec{\bar f}_{\rm excess}\;.
\end{eqnarray}
We also obtain the ``external" contribution to the bending moment
\begin{equation}
\label{eq:shapecon}
-\vec{\hat \gamma}^\alpha_*\times\vec{R}
=\vec{\omega}^\alpha_{(0)} -\vec{R}\times\vec{\gamma}^\alpha_{(0)}
=-\zeta_3 Q_{\rm excess}\varepsilon^{\alpha\beta}\vec{t}_\beta\;,
\end{equation}
which is only tangential.  Thus we can determine the bending-moment tensor to be
\begin{eqnarray}
\label{eq:bendex}
M^{\alpha\beta} &=& \lambda^{\alpha\beta}-\zeta_3 Q_{\rm excess} g^{\alpha\beta}\nonumber\\
&=& -\frac{1}{2}\left( \frac{\partial f_{}}{\partial H} +2\zeta_3 Q_{\rm excess} \right) g^{\alpha\beta}
	          -\frac{\partial f_{}}{\partial K}L^{\alpha\beta}\;.
\end{eqnarray}

We can give an explicit example of what the parameters $\zeta_1$, $\zeta_2$ and $\zeta_3$ could be, by assuming that the membrane can be modelled as consisting of a material that mimics an incompressible fluid. A fluid particle at distance $h$ from the membrane will then move to a distance ${\bar h}=h+\delta {\bar h}$ during a deformation in such a way that the fluid volume between the particle and the dividing surface is conserved. Mathematically we can write this criterion of volume conservation as (see Equation (\ref{eq:sqrtG}))
\begin{eqnarray}
0&=&\delta\left(\int_0^{\bar h}d h\;\sqrt{g}\left(1-2 h H +h^2 K\right)\right)\nonumber\\
&=&\delta\left(\sqrt{g}\left({\bar h}-{\bar h}^2 H+\frac{1}{3}{\bar h}^3 K\right)\right)\;.
\end{eqnarray}
If we isolate $\delta{\bar h}$ in the above expression and expand to second order in $h$ we find
\begin{equation}
\delta {\bar h}=-\left(h+h^2 H\right)\frac{\delta\sqrt{g}}{\sqrt{g}}+h^2\delta H+O\left((h^3\right)\;.
\end{equation}
Comparing with Equation (\ref{eq:gendeform}) we see that for this specific example we have
\begin{equation}
\zeta_1=\zeta_2=\zeta_3=\frac{1}{2}\;.\label{eq:zetahalf}
\end{equation}

An application of the considerations above can be found for a model of membrane activity proposed in \cite{manneville01} to explain experiments showing an increase in membrane shape fluctuations upon activation of certain proteins in the membrane. In this model the membrane proteins where activated by feeding them with the right energy source, which they then consumed to perform a specific task. While performing the task the proteins were assumed to change configuration or move nearby material in such a way that the proteins constantly pushed on their surroundings. Mathematically this was modelled as a distribution of forces added to the force-balance of the bulk fluids surrounding the membrane. In the language of this paper such a force distribution can be included by assigning it to the external force density $\vec{\bar f}_{\rm real,ext}$. A way to write down the model would then be
\begin{equation}
\vec{\bar f}_{\rm real,ext}(\vec{r})=
\int_{\rm M} d A\int d h\;F_{\rm act}(\uc^1,\uc^2,h)\vec{n}\delta^3\left(\vec{r}-(\vec{R}+h\vec{n})\right)\;,\label{eq:frealact}
\end{equation}
where the specific expression for $F_{\rm act}$ used in \cite{manneville01} was based on each protein contributing a force-dipole:
\begin{equation}
F_{\rm act}=\left(F_a (\rho^\uparrow-\rho^\downarrow)+2H F_a' (\rho^\uparrow+\rho^\downarrow)\right)\left[\delta\left(h-w^\uparrow\right)-\delta\left(h+w^\downarrow\right)\right]\;.\label{eq:MBRP}
\end{equation}
Here $\rho^\uparrow$ and $\rho^\downarrow$ represent area densities of the active proteins in the membrane, with $\uparrow$ and $\downarrow$ indicating the two possible orientations of an asymmetric transmembrane protein. $F_a$ and $F_a'$ are constants representing the strength of the active forces and their curvature dependence, $w^\uparrow$ and $w^\downarrow$ are constant lengths giving the distances from the membrane where the forces act. If we switch form the coordinate $\vec{r}$ to $(\uc^1,\uc^2,h)$ and use that $\vec{\bar f}_{\rm real,ext}$ has to be balanced by an equal but opposite $\vec{\bar f}_{\rm real}$ we can restate Equation (\ref{eq:frealact}) as
\begin{equation}
\vec{\bar f}_{\rm real}(\uc^1,\uc^2,h)=-\frac{\vec{n}F_{\rm act}(\uc^1,\uc^2,h)}{1-2 h H +h^2 K}\;.
\end{equation}
We can define a Gibbs system corresponding to this microscopic system by stating that in the Gibbs system the bulk fluids are free of external active forces, i.e. $\vec{\bar f}^\pm=0$. Then $\vec{\bar f}_{\rm excess}=\vec{\bar f}_{\rm real}$ for $h\ne 0$ and we get the simple expressions
\begin{eqnarray}
p_{\rm excess} &=&-\int d h\;h F_{\rm act}\;,\label{eq:pexc}\\
Q_{\rm excess} &=&-\int d h\;h^2 F_{\rm act}\;.\label{eq:Qexc}
\end{eqnarray}
The expression given for $\vec{\gamma}^\alpha_{(0)}$ in Equation (\ref{eq:gammaspec}), when supplemented by Equations (\ref{eq:zetahalf}), (\ref{eq:pexc}) and (\ref{eq:Qexc}), is exactly the expression for the active contribution to the membrane stress $\vec{T}^\alpha_{\rm act}$ derived in \cite{lomholt05c} in a different manner. Thus the considerations on the effect of external forces distributed in the transverse direction to the membrane developed in this paper provides an alternative point of view on the mechanics of this type of active membranes. We would like to point out that this kind of membrane activity is also an example of how the Gibbs description can be very useful. This was shown in \cite{lomholt05d} where the Gibbs description derived in \cite{lomholt05c} was used to calculate the fluctuation spectrum of an active quasi-spherical vesicle.

\section{Arbitrariness in the effective stresses}
\label{sec:arbit}
As mentioned already in Section~\ref{sec:thermodyn} in connection with the free-energy
variation, Equation~(\ref{eq:varfree}), there exists a certain degree of arbitrariness
in the definition of the variational contribution $\vec{S}^\alpha_{(0)}$ to the linear
stress $\vec{T}^\alpha$.  Related to that, there is also arbitrariness in the variational
contribution $\vec{Q}^\alpha_{(0)}$, defined in Equation~(\ref{eq:spatrot2}), to the angular
stress $\vec{\Omega}^\alpha$.  This issue has been mentioned in passing in some of the
earlier literature \cite{capovilla02}, and it plays an important role in the recent article \cite{guven06}, where it is proposed to find solutions of membrane shape equations by equating an effective membrane stress tensor with a null stress that represents exactly the arbitrariness discussed here. In this Section,
we discuss how to eliminate part of the arbitrariness the membrane stress tensor based on the understanding we have obtained of the effective mechanical quantities from the microscopic perspective.  Moreover we will also present a geometry-based interpretation of the remaining part of the arbitrariness. 

It is not difficult to see that the free-energy variation formulated in Equation~(\ref{eq:varfree})
is unchanged under a ``gauge" transformation of $\vec{\hat S}^\alpha$ of the following form:
\begin{equation}
\label{eq:limitarb}
\vec{{\hat S}'}^\alpha\cdot\delta\vec{R}
=\vec{\hat S}^\alpha\cdot \delta\vec{R} +\varepsilon^{\alpha\beta}\partial_\beta\left(\vec{\hat W}\cdot\delta\vec{R}\right)\;,
\end{equation}
where $\vec{\hat W}$ is an arbitrary vector-operator with the general form
\begin{equation}
\vec{\hat W}=\sum_{n=0}^\infty \vec{W}_{(n)}^{\beta_1\dots\beta_n}D_{\beta_1}\dots D_{\beta_n}\;.
\end{equation}

Equation~(\ref{eq:limitarb}) corresponds to a redefinition of the variational contribution to the
linear and angular stresses
\begin{eqnarray}
\vec{S'}^\alpha_{(0)}&=& \vec{S}^\alpha_{(0)} +\varepsilon^{\alpha\beta}\partial_\beta\vec{W}_{(0)}\;,\label{eq:stressarb}\\
\vec{Q'}^\alpha_{(0)}&=& \vec{Q}^\alpha_{(0)} -\varepsilon^{\alpha\beta}\partial_\beta\left(\vec{\hat W}\times\vec{R}\right)\nonumber\\
&=& \vec{Q}^\alpha_{(0)} +\varepsilon^{\alpha\beta}\partial_\beta\left(\vec{R}\times\vec{W}_{(0)}
  +\vec{W}_{(*)}\right)\;,\label{eq:angarb}
\end{eqnarray}
where
\begin{eqnarray}
\vec{W}_{(*)}&=&-\left(\vec{\hat W}-\vec{W}_{(0)}\right)\times\vec{R}\nonumber\\
&=&\sum_{n=1}^\infty D_{\beta_1}\dots D_{\beta_n}\vec{R}\times\vec{W}_{(n)}^{\beta_1\dots\beta_n}\;.
\end{eqnarray}
Note that there is more than sufficient freedom in the definition of $\vec{W}_{(*)}$ that allows
us to consider $\vec{W}_{(*)}$ as an arbitrary vector-function \footnote{Readers with knowledge of
de Rham Cohomology (see for instance \cite{warner83}) should be able to convince themselves that
at least locally (\ref{eq:limitarb}), (\ref{eq:stressarb}) and (\ref{eq:angarb}) constitute the
most general form of arbitrariness in $\vec{\hat S}^\alpha\cdot\delta\vec{R}$, $\vec{S}^\alpha_{(0)}$
and $\vec{Q}^\alpha_{(0)}$, provided that their divergences are well defined quantities.}.

The key to a physically meaningful elimination of at least some of the arbitrariness represented
by the above equations is given by the bending moments. There is indeed a physical requirement
on the bending moment, $\vec{N}^\alpha$, namely, $\vec{N}^\alpha \cdot\vec{n}=0$. This in turn
implies that it is natural to demand that the variational contribution, $\vec{N}^\alpha_{\rm rs}$, should satisfy $\vec{N}^\alpha_{\rm rs}\cdot\vec{n}=0$, and we will see that we have enough gauge freedom to ensure this requirement. In other words, we may use the arbitrariness in the stress definition to ensure that the physical requirement be fulfilled. Another condition is the condition that the tangential components of the variational bending moment ${M}^{\alpha\beta}_{\rm rs}=-\vec{N}^{\alpha}_{\rm rs}\cdot\vec{t}^\gamma\varepsilon_{\gamma}^{\phantom{\gamma}\beta}$ should be symmetric. This condition is a natural one when the bending moments may be described as derivatives of the free energy with respect to the symmetric curvature $K_{\alpha\beta}$.  But, it is not a general one, based on an understanding of a microscopic origin.

To see what that amounts to explicitly, we write out the ``gauge" transformation for
$\vec{N}^\alpha_{\rm rs}$, which reads as
\begin{equation}
\vec{N'}^\alpha_{\rm rs}
= \vec{N}^\alpha_{\rm rs} +\varepsilon^{\alpha\beta}\vec{t}_\beta\times\vec{W}_{(0)}
 + \varepsilon^{\alpha\beta}\partial_\beta\vec{W}_{(*)}\;.
\end{equation}
It should be clear that the second term on the right-hand side of this equation makes it quite
easy to choose the appropriate ``gauge" that satisfies the physical requirement.  It can be
checked easily that both $\vec{N'}^\alpha_{\rm rs}\cdot\vec{n}=0$ and a symmetric
${M'}^{\alpha\beta}_{\rm rs}$ can be achieved simultaneously by, for example,
the following gauge
\begin{eqnarray}
\vec{W}_{(0)} &=& \left(\vec{N}^\beta_{\rm rs}\cdot\vec{n}\right)\vec{t}_\beta
              -\left(\frac{1}{2}\vec{N}^{\alpha}_{\rm rs}\cdot\vec{t}_\alpha\right)\vec{n}\;,\\
\vec{W}_{(*)} &=&0 \;.
\end{eqnarray}

It turns out that the two conditions mentioned above do not eliminate the arbitrariness,
or ``gauge freedom," completely.  The remaining gauge freedom is described by a transformation
\begin{eqnarray}
\vec{W}_{(0)}
&=& \frac{1}{2}\varepsilon^{\alpha\beta}\left(\partial_\alpha\vec{\Lambda}\cdot\vec{t}_\beta\right)\vec{n}
  -\varepsilon^{\alpha\beta}\left(\partial_\alpha\vec{\Lambda}\cdot\vec{n}\right)\vec{t}_\beta\;,\label{eq:Varb}\\
\vec{W}_{(*)}
&=&\vec{\Lambda}\;, \label{eq:Warb}
\end{eqnarray}
where $\vec{\Lambda}$ is an arbitrary vector-function.  It connects a sequence of
$\vec{N}^\alpha_{\rm rs}$'s and ${M}^{\alpha\beta}_{\rm rs}$'s, which all satisfy the two
conditions.  In terms of $\vec{N}^\alpha_{\rm rs}$ and ${M}^{\alpha\beta}_{\rm rs}$, the
transformation reads as
\begin{equation}
\vec{N'}^\alpha_{\rm rs}
= \vec{N}^\alpha_{\rm rs} +\left( \varepsilon^{\alpha\beta}\delta^\gamma_{\phantom{\gamma}\delta}
                             +\frac{1}{2}\delta^\alpha_{\phantom{\delta}\delta}\varepsilon^{\beta\gamma}
			\right) \left(\partial_\beta\vec{\Lambda}\cdot\vec{t}_\gamma\right) \vec{t}^\delta\;,
\end{equation}
or equivalently,
\begin{equation}
{M'}^{\alpha\beta}_{\rm rs}
= M^{\alpha\beta}_{\rm rs} -\frac{1}{2}\left(\partial^{\alpha}\vec{\Lambda}\cdot\vec{t}^{\beta}+\partial^{\beta}\vec{\Lambda}\cdot\vec{t}^{\alpha}\right)
 + \partial^{\gamma}\vec{\Lambda}\cdot\vec{t}_{\gamma}g^{\alpha\beta}\;.
\end{equation}

\subsection{Arbitrariness from the Codazzi-Mainardi equations and theorema egregium}
No more obvious conditions that are physically meaningful can be imposed on the effective
stresses derived variationally to eliminate the remaining arbitrariness, represented
by the vector function $\vec{\Lambda}$.  In this Subsection, we develop a geometrical
interpretation of $\vec{\Lambda}$.  This interpretation may help a user of the membrane
mechanics make the most judicious choice, or gauge, in practice.  

The geometrical interpretation is derived from the Codazzi-Mainardi equations, given in
Equation~(\ref{eq:GCM2}), and theorema egregium, formulated as in Equation~(\ref{eq:GCM1}).  To describe
qualitatively why the arbitrariness should be related to these equations, we recall first
that a surface is uniquely defined, modulo rotations, translations and reflections, by
the two tensors $g_{\alpha\beta}$ and $K_{\alpha\beta}$ \cite{spivak70}.  The metric tensor
$g_{\alpha\beta}$ describes the local extension/compression of the surface, the type of
deformation from which the linear stress originates; the curvature tensor $K_{\alpha\beta}$
describes the local bending of the surface, apparently the type of deformation that is
associated with the angular stress or the bending moments.  Were $g_{\alpha\beta}$ and
$K_{\alpha\beta}$ independent of each other, we would be able to derive $\sigma^{\alpha\beta}$
and $ \lambda^{\alpha\beta}$, generally and uniquely, from Equation~(\ref{eq:metriccurvform})
as functional derivatives of the free energy, and then determine the stresses and bending
moments uniquely, as in Equation~(\ref{eq:Ttex}) and Equation~(\ref{eq:varcon}).  $g_{\alpha\beta}$
and $K_{\alpha\beta}$ can not, however, be varied independently, due to the fact that they are tied together precisely
through the Codazzi-Mainardi equations and theorema egregium.  In other words, general
bending and stretching of a surface can not be decoupled completely.  Thus, the geometrical
constraints should be related to the arbitrariness of the stress tensor.

To show the connection more explicitly, we note first that the Riemann tensor
$\mathcal{R}_{\alpha\beta\gamma\delta}$ for two-dimensional geometrical manifolds can be written
as
\begin{equation}
\mathcal{R}_{\alpha\beta\gamma\delta}=\frac{\mathcal{R}}{2}\varepsilon_{\alpha\beta}\varepsilon_{\gamma\delta}\;.
\end{equation}
The full non-trivial content of theorema egregium is thus captured by Equation~(\ref{eq:GaussvsRie}):
\begin{equation}
\mathcal{R}=2K\;.\nonumber
\end{equation}
We then consider a ``contribution'' to the free-energy that reads as
\begin{equation}
f=\omega_{\rm n}\left(\mathcal{R}/2-K\right)\;.
\end{equation}
This term is identically zero, and, therefore, makes no real contribution to the total free energy.
The contributions to the linear and angular stresses derived directly from the variation of this zero free energy, using the definitions in section \ref{sec:geom} and equations (\ref{eq:metriccurvform}), (\ref{eq:Ttex}) and (\ref{eq:Tnex}), are, however, not zero, although their divergences must be zero.

To calculate the variation, we first evaluate the variation of the Christoffel symbols, which are given by
\begin{equation}
\delta\Gamma^\gamma_{\alpha\beta}
=\frac{1}{2} g^{\gamma\delta} \left( D_\beta\delta g_{\delta\alpha}+D_\alpha\delta g_{\delta\beta}
                               - D_\delta\delta g_{\alpha\beta}\right)\;.
\end{equation}
From this we can then derive, based on Equation~(\ref{eq:Riedef}), 
\begin{equation}
\delta\mathcal{R}^\gamma_{\phantom{\gamma}\alpha\delta\beta}
= D_\delta\left(\delta\Gamma^\gamma_{\alpha\beta}\right)
-D_\beta\left(\delta\Gamma^\gamma_{\alpha\delta}\right)\;, 
\end{equation}
and in turn
\begin{equation}
\delta\mathcal{R}
= \left( -\frac{\mathcal{R}}{2}g^{\alpha\beta} +D^\alpha D^\beta 
       -g^{\alpha\beta}D^\gamma D_\gamma \right) \delta g_{\alpha\beta}\;.
\end{equation}
Using the above result, we finally arrive at a formal expression of the free-energy variation:
\begin{eqnarray}
\frac{1}{\sqrt{g}}\delta\left(\sqrt{g}f\right)&=& \frac{1}{2}\sigma^{\alpha\beta}\delta g_{\alpha\beta}-\lambda^{\alpha\beta}\delta K_{\alpha\beta}\nonumber\\
 && +\frac{1}{2}D_\gamma\big[ \big( -\omega_{\rm n}g^{\alpha\beta}D^\gamma   + \omega_{\rm n}g^{\gamma\alpha}D^\beta\nonumber\\
&& \phantom{+\frac{1}{2}D_\gamma\big[ \big( } +D^\gamma\omega_{\rm n} g^{\alpha\beta}
          -D^\alpha\omega_{\rm n}g^{\gamma\beta} \big)\delta g_{\alpha\beta} \big]\,, \label{eq:Cod1var}
\end{eqnarray}
where
\begin{eqnarray}
\sigma^{\alpha\beta} &=& \omega_{\rm n}\left( 2K-\frac{\mathcal{R}}{2} \right)
                    + D^\alpha D^\beta\omega_{\rm n} -D^\gamma D_\gamma\omega_{\rm n} g^{\alpha\beta}\;,\\
\lambda^{\alpha\beta} &=& \omega_{\rm n}L^{\alpha\beta}\;.
\end{eqnarray}
Note that the boundary term inside the square brackets in the last line of Equation~(\ref{eq:Cod1var}) is
invariant under both rigid translations and rotations, and will not, therefore, contribute to the
stress tensor or the bending moment. 

We can now use Equation~(\ref{eq:deltag2}) and Equation~(\ref{eq:deltaK2}) to read off from Equation~(\ref{eq:Cod1var}) the contributions to the stresses, which are
\begin{eqnarray}
M^{\alpha\beta}_{\rm rs} &=& \lambda^{\alpha\beta}=\omega_{\rm n}L^{\alpha\beta}\;,\label{eq:Mgaussarb}\\
\vec{S}^\alpha_{(0)}\cdot\vec{n} &=& D_\beta\lambda^{\beta\alpha}=L^{\alpha\beta}\partial_\beta\omega_{\rm n}\;,\\
\vec{S}^{\alpha}_{(0)}\cdot\vec{t}^\beta 
&=& \sigma^{\alpha\beta}-\lambda^{\alpha\gamma}K_{\gamma}^{\phantom{\gamma}\beta}
  =D^\alpha D^\beta\omega_{\rm n}-g^{\alpha\beta}D^\gamma D_{\gamma}\omega_{\rm n}\label{eq:Sgaussarb}\;.
\end{eqnarray}

It is conceptually obvious that the above contributions necessarily imply a certain degree of
arbitrariness in the definitions of the stresses, since the starting point is a zero free
energy.  That this is the case can also be seen, if we make a gauge transformation of the
form given in Equation~(\ref{eq:Varb}) and Equation~(\ref{eq:Warb}) with $\vec{\Lambda}=\omega_{\rm n}\vec{n}$:
the stresses become zero in the new gauge.  

The Codazzi-Mainardi equations are related to the tangential components of the vector function 
$\vec{\Lambda}$.  The proof is similar to what has been given above.  We give only a brief
presentation.  The non-trivial content of the equations is captured by
\begin{equation}
D_\beta K^\beta_{\phantom{\beta}\alpha}=D_\alpha K^\beta_{\phantom{\beta}\beta}\;.
\end{equation}
We consider, therefore, the following zero free energy
\begin{equation}
f=\omega^\alpha\left(D_\beta K^\beta_{\phantom{\beta}\alpha}-D_\alpha K^\beta_{\phantom{\beta}\beta}\right)\;.
\end{equation}
Performing the variation in the same way as before, we get
\begin{eqnarray}
M^{\alpha\beta}_{\rm rs}&=&\frac{1}{2}\left(D^{\alpha}w^{\beta}+D^{\beta}w^{\alpha}\right)-D_\gamma\omega^\gamma g^{\alpha\beta}\;,\\
\vec{S}^\alpha\cdot\vec{n}&=&D_\beta M^{\beta\alpha}\;,\\
\vec{S}^\alpha_{(0)}\cdot\vec{t}^\beta
&=&K^{\alpha}_{\phantom{\alpha}\gamma}D^\beta w^\gamma
  -K^{\gamma\delta}D_\gamma w_\delta g^{\alpha\beta} \nonumber\\
&&+\frac{1}{2}K^{\beta}_{\phantom{\beta}\gamma} \left( D^\alpha w^\gamma-D^\gamma w^\alpha \right)
 -w^\gamma D_\gamma L^{\alpha\beta}\;.
\end{eqnarray}
A useful mathematical fact for doing these calculations is that in two dimensions, the following
identity holds for any tensor $A^{\alpha\beta}$ 
\begin{equation}
\varepsilon^{\alpha\gamma}A_{\gamma\delta}\varepsilon^{\delta\beta}
= A^{\beta\alpha}-A^{\gamma}_{\phantom{\gamma}\gamma}g^{\alpha\beta}\;.
\end{equation}
These stresses in fact represent the rest of the remaining gauge freedom, since they become
zero under the gauge transformation defined by $\vec{\Lambda}=\omega^\alpha\vec{t}_\alpha$.

\section{Conclusion}\label{sec:concl}
Canonically, mechanics of a fluid membrane is effectively described as the mechanics of an
infinitely thin surface.  In other words, the effective description must reflect the physics
that is associated with the finite, albeit microscopic, thickness of the membrane.  Moreover,
the effective description is formulated either in terms of the concepts of mechanical deformation
and stresses, or in terms of mechanical free-energy functions.  The connection between these
two descriptions has been a subject of discussion in the existing literature
\cite{evans80,kralchevsky94,capovilla02}.  A number of issues in this context have not, however,
been addressed clearly and fully, or at all.  In particular, they include the issue of what
the microscopic origins are of the effective mechanical stresses, as well as the issue of what
the connection is between the two descriptions when external forces act on the membrane system
under consideration.

It is the main purpose of this paper to address these issues, in a way that both casts the
different existing works in a coherent framework and extends the current framework.  To this
end, we have approached both of the effective descriptions from a microscopic perspective,
where the fluid membrane is treated as a microscopically thin layer, with highly inhomogeneous
material and force distributions in its transverse direction, and have demonstrated unambiguously
how the effective descriptions arise from the microscopic perspective, i.e., the microscopic
origins of the surface mechanical stresses.  Moreover and more importantly, we have, facilitated
by the microscopic perspective, established a general connection between the mechanical-stress
based and the free-energy based descriptions.  Naturally, in doing so we have recovered the
canonical connection already established in the existing literature, but we have also addressed
a specific issue that is important to practical applications of the energy-based description of
membrane mechanics: the issue of arbitrariness involved in identifying the mechanical stresses
from a given free energy.  We hope that we have provided more insight into the issue and that
the understanding will facilitate the utility of both of the descriptions of membrane mechanics.
Furthermore, we have worked out the connection for situations where there act external forces
on the system.  This is a non-trivial extension to the existing theories.  We expect that it
will be relevant and useful to further studies of membrane mechanics, both theoretically and
experimentally.

\subsection{acknowledgments}
MAL acknowledges the financial support from University of Southern Denmark and the Villum Kann Rasmussen Foundation. LM would like to thank the Danish National Research Foundation for its
financial support in the form of a long-term operating grant awarded to The MEMPHYS-Center
for Biomembrane Physics. The authors are grateful to Tove Nyberg for her technical assistance
on the figures.
%\subsection{acknowledgments}

\appendix

\section{Differential geometry of surfaces}
\label{sec:geom}
In this appendix we briefly review the mathematical language of differential geometry of two dimensional surfaces. A more comprehensive introduction can be found in Refs. \cite{kreyszig91,spivak70}, for example.

The shape of a two-dimensional surface is represented by a space-vector function $\vec{R}=\vec{R}(\uc^1,\uc^2)$.  The variables $\uc^1$ and $\uc^2$ are internal coordinates corresponding to a parametrization of the surface.  At each point on the surface a basis for three dimensional vectors can be established.  Two of them are tangential vectors, defined as
\begin{equation}
\vec{t}_\alpha \equiv \partial_\alpha\vec{R}\equiv \frac{\partial \vec{R}}{\partial \uc^\alpha}\;,
\end{equation}
where $\alpha=1,2$, and the third is a unit vector normal to the surface, given by 
\begin{equation}
\label{eq:normaldef}
\vec{n} \equiv \frac{\vec{t}_1\times\vec{t}_2}{|\vec{t}_1\times\vec{t}_2|}\ .
\end{equation}

Local geometry of the surface is characterized by two surface tensors, the metric tensor and the curvature tensor.  The local metric tensor is defined by
\begin{equation}
g_{\alpha\beta} \equiv \vec{t}_\alpha\cdot\vec{t}_\beta\ .
\end{equation}
It has an inverse, $g^{\alpha\beta}$, which satisfies, by definition,
\begin{equation}
g^{\alpha\beta}g_{\beta\gamma} = \delta^\alpha_\gamma\;,
\end{equation}
where $\delta^\alpha_\gamma$ is the Kronecker delta and where the repeated Greek superscript-subscript indices imply summation following the Einstein summation convention.  The metric tensor and its inverse are used to raise and lower Greek indices as in the following example:
\begin{equation}
\vec{t}^\alpha=g^{\alpha\beta}\vec{t}_\beta\;,\quad \vec{t}_\alpha=g_{\alpha\beta}\vec{t}^\beta\;.
\end{equation}
The curvature tensor $K_{\alpha\beta}$ is defined via the second derivatives of the surface shape function,
\begin{equation}
K_{\alpha\beta}\equiv\vec{n}\cdot\partial_\alpha\partial_\beta\vec{R}\;.
\end{equation}
From it the scalar mean curvature $H$ and the Gaussian curvature $K$ can be obtained:
\begin{eqnarray}
H&=&\frac{1}{2}g^{\alpha\beta}K_{\alpha\beta}\ ,\\
K&=&\det g^{\alpha\beta}K_{\beta\gamma}\ .
\end{eqnarray}

Two other tensors will also be introduced here,
\begin{equation}
\varepsilon_{\alpha \beta} \equiv \epsilon_{\alpha \beta} \sqrt{g} \;, \quad \quad \quad
\varepsilon^{\alpha \beta} \equiv \epsilon^{\alpha \beta}/ \sqrt{g} \;,
\end{equation}
where $\epsilon_{\alpha \beta} = \epsilon^{\alpha \beta}$ with $\epsilon_{11}= \epsilon_{22}= 0$ and $\epsilon_{12} = - \epsilon_{21} = 1$ are tensor densities, and $g=\det g_{\alpha\beta}$ is the determinant of the metric tensor.

Expressions of covariant/contravariant differentiations of vector and tensor functions defined on the surfaces are facilitated by the use of the Christoffel symbols, $\Gamma^\gamma_{\alpha\beta}$. One instance, which will become particularly useful later, is the covariant differentiation of a surface vector function, $\vec{w}=w^\alpha\vec{t}_\alpha$, given by
\begin{equation}
D_\alpha w^\beta = \partial_\alpha w^\beta + w^\gamma \Gamma^\beta_{\gamma\alpha}\;.
\end{equation}
The Christoffel symbols can also be defined as certain combinations of the derivatives of the metric tensor, namely,
\begin{equation}\label{eq:Chrdef}
{\Gamma}^\gamma_{\alpha\beta}=\frac{1}{2}g^{\gamma\delta}\left(\partial_\alpha{g}_{\beta\delta}
                            +\partial_\beta{g}_{\delta\alpha}-\partial_\delta{g}_{\beta\alpha}\right)\;.
\end{equation}
It follows that the covariant divergence of $w^\alpha$ can be written as 
\begin{equation}
D_\alpha w^\alpha=\frac{1}{\sqrt{g}}\partial_\alpha \left(\sqrt{g}w^\alpha\right)\;.
\end{equation}

The area of a local differential element of the surface is given by
\begin{equation}
dA=\sqrt{g}d\uc^1d\uc^2\;,
\end{equation}
an expression which will be repeatedly used in surface integrals.  

Two equations that are frequently used in this paper are the Gauss formula
\begin{equation}
D_\alpha\vec{t}_\beta=K_{\alpha\beta}\vec{n}\;,\label{eq:GW1}
\end{equation}
and the Weingarten equations
\begin{equation}
D_\alpha\vec{n}=-K_{\alpha\beta}\vec{t}^\beta\;.\label{eq:GW2}
\end{equation}
Equation~(\ref{eq:GW1}) follows from the definition of the curvature tensor and the definition of covariant differentiation, and Equation~(\ref{eq:GW2}) is obtained by differentiating $\vec{n}\cdot\vec{n}=1$ and $\vec{n}\cdot\vec{t}_\alpha=0$ and solving for $D_\alpha\vec{n}$.

We will also from time to time weed both the Codazzi-Mainardi equations
\begin{equation}
D_\alpha K_{\beta\gamma}-D_\beta K_{\alpha\gamma}=0\;,\label{eq:GCM2}
\end{equation}
and an equation related to the famous {\it theorema egregium} by Gauss
\begin{equation}
\mathcal{R}_{\alpha\beta\gamma\delta}-K_{\alpha\gamma}K_{\beta\delta}+K_{\alpha\delta}K_{\beta\gamma}=0\;,
\label{eq:GCM1}
\end{equation}
where the Riemannian curvature $\mathcal{R}^\gamma_{\phantom{\gamma}\alpha\delta\beta}$ can be calculated from the Christoffel symbols by
\begin{equation}\label{eq:Riedef}
\mathcal{R}^\gamma_{\phantom{\gamma}\alpha\delta\beta} 
= \partial_\delta\Gamma^\gamma_{\alpha\beta}-\partial_\beta\Gamma^\gamma_{\alpha\delta}
  +\Gamma^\gamma_{\epsilon\delta}\Gamma^\epsilon_{\alpha\beta}
  -\Gamma^\gamma_{\epsilon\beta}\Gamma^\epsilon_{\alpha\delta}\;.
\end{equation}

Another way to define the Riemannian curvature is through covariant differentiation of an arbitrary vector $v^\gamma$:
\begin{equation}\label{eq:Rie2def}
\left(D_\alpha D_\beta-D_\beta D_\alpha\right)v^\gamma
=\mathcal{R}^\gamma_{\phantom{\gamma}\delta\alpha\beta}v^\delta\;.
\end{equation}
In other words $\mathcal{R}^\gamma_{\phantom{\gamma}\alpha\delta\beta}$ measures the degree to which covariant differentiations commute.

Contracting the Riemannian curvature tensor we get the Ricci curvature tensor
\begin{equation}
\mathcal{R}_{\alpha\beta}=\mathcal{R}^\gamma_{\phantom{\gamma}\alpha\gamma\beta}\ ,
\end{equation}
and further taking the trace we get the Ricci scalar
\begin{equation}
\mathcal{R}=g^{\alpha\beta}\mathcal{R}_{\alpha\beta}\ .
\end{equation}
If we contract Equation~(\ref{eq:GCM1}) twice with the metric and use the following identity
\begin{equation}
K_{\alpha\gamma}K^{\gamma}_{\phantom{\gamma}\beta}=2HK_{\alpha\beta}-Kg_{\alpha\beta}\;,
\end{equation}
we see that the Ricci scalar and the Gaussian curvature are intimately related as
\begin{equation}
\mathcal{R}=2K\;.\label{eq:GaussvsRie}
\end{equation}
This is, in fact, one way to state {\em theorema egregium}.

Theorema egregium and the Codazzi-Mainardi equations relate the intrinsic geometrical properties of the surface, i.e. those associated with the metric $g_{\alpha\beta}$, to the extrinsic properties associated with $K_{\alpha\beta}$.  In the main part of the paper it is shown how these relations lead to some arbitrariness in the definition of stresses in membranes.

There is a symmetric tensor which will be used sufficiently often
to merit a symbol on its own.  It is defined as
\begin{equation}
L^{\beta\gamma}\equiv \varepsilon^{\beta\delta}\varepsilon^{\gamma\epsilon}K_{\delta\epsilon}
= 2Hg^{\beta\gamma}-K^{\beta\gamma}\;.
\end{equation}
This tensor is proportional to the inverse of the extrinsic curvature tensor (when the inverse exists)
\begin{equation}
L^{\beta\gamma}K_{\gamma\delta}=K\delta^\beta_{\delta}\;.
\end{equation}
It can be seen that
\begin{equation}
D_\beta L^{\beta\gamma}=0\;,
\end{equation}
which follows from
\begin{equation}
D_\gamma\varepsilon^{\alpha\beta}=0\;,
\end{equation}
and from Equation~(\ref{eq:GCM2}).

\section{The form of the membrane stress tensor}\label{sec:form}

\begin{figure}
\begin{center}
\resizebox{6cm}{!}{\includegraphics{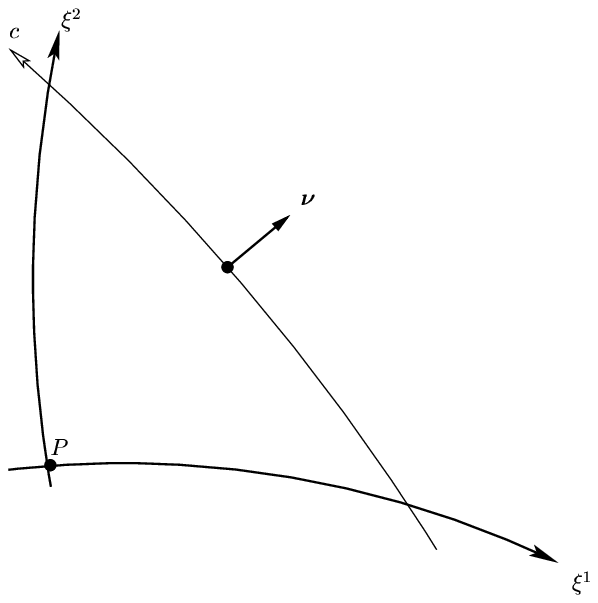}}
\end{center}
\caption{\label{fig:triangle}A triangle bounded by coordinate curves and the curve $c$.}
\end{figure}

In this appendix we will establish the form of the stress tensor given in Equation~(\ref{eq:nudep}).

To establish Equation~(\ref{eq:nudep}) let us consider a small triangle in the membrane surface,
as illustrated in Figure~\ref{fig:triangle}.  The triangle is bounded by a curve $c$ and two
other coordinate curves crossing each other at a point $P$.  The outward pointing unit normal
to the $\uc^1$ coordinate curve is $\vec{\nu}^{(1)}=\vec{t}_1\times\vec{n}/\sqrt{g_{11}}$. 
The edge of the triangle formed by this curve has a length
$ds_{(1)}=\sqrt{g_{11}}d\uc^1=\sqrt{g^{22}}\nu_2ds$, where $d\uc^1$ is the coordinate length
of the curve and $ds$ is the length of the part of curve $c$ that contributes to the triangle. 
Similarly, the outward pointing unit normal to the other coordinate curve is
$\vec{\nu}^{(2)}={\vec{n}\times\vec{t}_2/\sqrt{g_{22}}}$, and the length of the relevant part
of the curve is $ds_{(2)}=\sqrt{g_{22}}d\uc^2=\sqrt{g^{11}}\nu_1ds$. If we define
\begin{equation}\label{eq:2Dstressdef}
\vec{T}^1=-\sqrt{g^{11}}\vec{T}_{(\vec{\nu}^{(2)})}\;,
\vec{T}^2=-\sqrt{g^{22}}\vec{T}_{(\vec{\nu}^{(1)})}\;,
\end{equation}
then the force on the $\uc^1$ coordinate curve is $(-\vec{T}^2/\sqrt{g^{22}})ds_{(1)}$, and
the force on the $\uc^2$ curve is $(-\vec{T}^1/\sqrt{g^{11}})ds_{(2)}$.  The sum of the forces
on the triangle is
\begin{equation}
\vec{T}_{(\vec{\nu})}ds -\vec{T}^2 \frac{ds_{(1)}}{\sqrt{g^{22}}}
  -\vec{T}^1 \frac{ds_{(2)}}{\sqrt{g^{11}}} = \left(\vec{T}_{(\vec{\nu})}-\vec{T}^2\nu_2-\vec{T}^1\nu_1\right)ds \;.
\end{equation}
The sum of the forces should decrease at least proportionally to the area of the triangle,
as the triangle is reduced by moving curve $c$ closer and closer to Point $P$, in order 
that a force density of normal magnitude exist.  In other words,
$\left(\vec{T}_{(\vec{\nu})}-\vec{T}^2\nu_2-\vec{T}^1\nu_1\right)$ should approach zero
as $ds$ approaches zero, from which Equation~(\ref{eq:nudep}) follows.

\section*{References}
\bibliography{mycites}

\end{document}